\documentclass[traditabstract]{aa}
\usepackage{graphicx}
\usepackage{txfonts}
\usepackage{hyperref}
\usepackage{xcolor}
\usepackage{amsmath}
\usepackage{natbib}
\usepackage{lscape}  
\usepackage{multirow}

\hypersetup{
     colorlinks   = True,
     citecolor    = blue,
     linkcolor = blue,
     urlcolor = blue
}

\begin{document}

  \title{The VIMOS Ultra-Deep Survey:  A major merger origin for the high fraction of galaxies at $2<z<6$ with two bright clumps  \thanks{Based on data obtained with the European 
          Southern Observatory Very Large Telescope, Paranal, Chile, under Large
          Program 185.A--0791. }}

\author{
B. Ribeiro \inst{1}
\and O.~Le F\`evre\inst{1}
\and P.~Cassata\inst{2}
\and B.~Garilli\inst{3}
\and B.~C.~Lemaux \inst{1,4}
\and D.~Maccagni\inst{3}
\and D. Schaerer\inst{5,6}
\and L. A. M.~Tasca\inst{1}
\and G.~Zamorani \inst{7}
\and E.~Zucca\inst{7}
\and R.~Amor\'in\inst{8,9}
\and S.~Bardelli\inst{7}
\and N.P.~Hathi\inst{1,10}
\and A.~Koekemoer\inst{10}
\and J.~Pforr\inst{1,11}
}

\institute{Aix Marseille Univ, CNRS, LAM, Laboratoire d'Astrophysique de Marseille, Marseille, France
\and
Instituto de Fisica y Astronom\'ia, Facultad de Ciencias, Universidad de Valpara\'iso, Gran Breta$\rm{\tilde{n}}$a 1111, Playa Ancha, Valpara\'iso, Chile
\and
INAF--IASF Milano, via Bassini 15, I--20133, Milano, Italy
\and
Department of Physics, University of California, Davis, One Shields Ave., Davis, CA 95616, USA
\and
Geneva Observatory, University of Geneva, ch. des Maillettes 51, CH-1290 Versoix, Switzerland
\and
Institut de Recherche en Astrophysique et Plan\'etologie - IRAP, CNRS, Universit\'e de Toulouse, UPS-OMP, 14, avenue E. Belin, F31400
Toulouse, France
\and
INAF--Osservatorio Astronomico di Bologna, via Ranzani, 1 - 40127, Bologna 
\and
Cavendish Laboratory, University of Cambridge, 19 JJ Thomson Avenue, Cambridge, CB3 0HE, UK
\and
 Kavli Institute for Cosmology, University of Cambridge, Madingley Road, Cambridge CB3 0HA, UK 
\and
Space Telescope Science Institute, 3700 San Martin Drive, Baltimore, MD 21218, USA
\and
Scientific Support Office, Directorate of Science and Robotic Exploration, ESA-ESTEC, Keplerlaan 1, 2201 AZ Noordwijk, The Netherlands
 \\ \\
             \email{bruno.ribeiro@lam.fr}
}

   \date{}

 
  \abstract{The properties of stellar clumps in star-forming galaxies and their evolution over the redshift range $2\lesssim z \lesssim 6$ are presented and discussed in the context of the build-up of massive galaxies at early cosmic times. We focused on galaxies with spectroscopic redshifts from the VIMOS Ultra Deep Survey (VUDS) and stellar masses $\log_{10}({M_\star/M_\odot}) > -0.204\times(z-4.5)+9.35$.  We analyzed HST-ACS images to identify clumps within a 20 kpc radius using a method taking into account differential surface brightness dimming and luminosity evolution with redshift. We find that the population of galaxies with more than one clump is dominated by galaxies with two clumps, representing $\sim$21-25\%  of the population, while the fraction of galaxies with three, or four and more, clumps  is 8-11 and 7-9\%, respectively.  The  fraction of clumpy galaxies is in the range $\sim35-55\%$ over $2<z<6$,  increasing at higher redshifts, indicating that the fraction of irregular galaxies remains high up to the highest redshifts. The large and bright clumps (M$_{star}\sim10^9$  up to $\sim10^{10}$M$_\sun$) are found to reside predominantly in galaxies with two clumps. Smaller and lower luminosity clumps (M$_{star}<10^{9}$M$_\sun$) are found in galaxies with three clumps or more. We interpret these results as evidence for two different modes of clump formation working in parallel.  The small low luminosity clumps are likely the result of disk fragmentation, with violent disk instabilities (VDI) forming several long-lived clumps in-situ. as suggested from simulations. A fraction of these clumps is also likely coming from minor mergers as confirmed from spectroscopy in several cases. The clumps in the dominating population of galaxies with two clumps are significantly more massive and have properties akin to those in galaxy pairs undergoing massive merging observed at similar redshifts; they appear as more massive than the most massive clumps observed in numerical simulations of disks with VDI.   We infer from these properties that the bright and large clumps are most likely the result of major mergers bringing-in ex-situ matter onto a galaxy, and we derive a high major merger fraction of  $\sim$20\%. The diversity of clump properties therefore suggests that the assembly of star-forming galaxies at z$\sim 2-6$ proceeds from several different dissipative processes including an important contribution from major and minor mergers.}


\authorrunning{Bruno Ribeiro et al.}
\titlerunning{A high fraction of galaxies at $2<z<6$ with two bright clumps}

   \maketitle
%

\section{Introduction}\label{sec:intro}

The morphology of galaxies is a visible outcome of all processes at work when galaxies are assembling along cosmic time \citep[e.g.,][]{conselice2014}. While at lower redshifts galaxies are predominantly regular and symmetric in the form of elliptical or spiral galaxies, the situation at redshifts higher than the peak in star formation z$\simeq$1.5-2 is very different with galaxies mostly having irregular shapes \citep[e.g.,][]{buitrago2013,mortlock2013,huertas-company2015}. 

Quantitative morphology allows us to follow the evolution of the two main components of  galaxies, disks and bulges, and studies have focused on understanding the evolutionary paths of these fundamental components \citep[e.g.,][]{bruce2012,bruce2014a,bruce2014b,lang2014,tasca2014b,margalef-bentabol2016}. 
Observational studies and numerical simulations seem to support a picture where the original baryonic matter collapsing in a dark matter halo conserves its angular momentum which leads to the formation a rotation supported proto-disk. In this scenario, violent disk instabilities (VDI) develop and then form large star-forming clumps. These large clumps may then migrate to the disk center to form the proto-bulges of spiral galaxies \citep[e.g.,][]{genzel2008,elmegreen2008,elmegreen2009,dekel2013,perez2013,bournaud2014,bournaud2016}. Alternatively, considerable effort has been invested in studying the development of elliptical galaxies along cosmic time. On this scenario, bulges would form from violent dissipative events, going through a stage of compact quiescent objects at z$\sim$2. Such objects would then grow into massive ellipticals after star formation quenching and a succession of merging events \citep[e.g.,][]{hopkins2008,oser2012,Lopez2012,porter2014}.
From a theory and simulations perspective, two major physical processes are therefore thought to play a key role in the evolution of different types of galaxies: gas accretion mostly in cold form and flowing along the cosmic web and/or smoothly distributed {\citep[e.g.,][]{dekel2009,rodriguez-gomez2016}}, and major or minor merging {\citep[e.g.,][]{hopkins2008,diMatteo2008}}, each leading to different morphological signatures.

An important  morphological property related to these processes is the number of large clumps of gas and stars present within a galaxy. Clumps may have either an in-situ or an ex-situ formation history depending on which physical processes are at play \citep[e.g.,][]{shibuya2016}.  Clumps forming from disk instabilities are mostly the result of in-situ dynamical processes, while major and minor merging events are driving clumps of ex-situ material into a galaxy in assembly phase. The properties and fate of these clumps have been extensively studied in the literature both from observations \citep[e.g.,][]{elmegreen2004,elmegreen2006,elmegreen2007,elmegreen2013,genzel2008,bournaud2008,wuyts2012,guo2012} and in hydrodynamical simulations \citep[e.g.,][]{bournaud2007,dekel2009,genel2012,bournaud2014,mandelker2014,tamburello2015,oklopcic2016}. Both report clumps with typical stellar masses of $10^{7}-10^{9}M_\odot$ and with short to long lifetimes of $\lesssim1$ Gyr. Some studies suggest that short-lived clumps are destroyed by internal feedback before migrating to the center of the galaxy \citep[e.g.,][]{genel2012}.

The measurement of clumps at redshifts beyond the peak of cosmic star formation is  notoriously difficult. The identification of clumpy galaxies in the dominant population of irregular galaxies at high redshift  dates back to the first deep \emph{Hubble Space Telescope} (HST) images \citep[e.g.,][]{williams1996}. Subsequent studies at $z \sim 1-3$  revealed that clumpy galaxies are more numerous than in the local Universe 
\citep[e.g.,][]{abraham1996,vanderbergh1996,giavalisco1996,elmegreen2004,elmegreen2006,elmegreen2007,elmegreen2008,elmegreen2009,elmegreen2013,kubo2013,kubo2016,glazebrook2013,tadaki2014,murata2014,guo2015,garland2015,bournaud2016}. The fraction of clumpy galaxies was studied from z=0 up to the most distant galaxies identified today at z$\sim$10 \citep{ravindranath2006,guo2012,guo2015,shibuya2016}. In the sample of galaxies with $0.5<z<3$ in the CANDELS survey, \citet{guo2015} find that galaxies with low mass $log(M_{star}/M_{\sun})<9.8$ have a high fraction of off-center clumps $f_{clumpy}\sim 60$\%  constant over the observed redshift range, while for intermediate and massive galaxies $f_{clumpy}$ decreases from $\sim$40\% at z$\sim$3 to $\sim$15\% at z$\sim$0.5. Combining deep HST imaging from the 3D-HST, CANDELS, HUDF and HFF surveys \citet{shibuya2016} claim that $f_{clumpy}^{UV}$ follows an evolution similar to that of the the star formation rate density \citep[e.g.,][]{madau2014}, increasing from z$\simeq$8 to z$\simeq 1-3$ and decreasing from z$\simeq$1 to z$\simeq$0.  In comparing these observed trends with the predictions of simulations, \citet{guo2015} conclude that VDI are likely responsible for $f_{clumpy}$ at high mass and  that minor mergers are a viable explanation for $f_{clumpy}$ at intermediate  mass for z$<$1.5, while \citet{shibuya2016} conclude that VDI is the main origin of all clumps. Both these studies exclude major mergers as a possible contribution to the evolution of the fraction of clumps. This is somewhat surprising as the major merger fraction for star-forming galaxies is observed to increase up to $\sim 20$\% at z$\simeq 1-2$ \citep[e.g.,][]{lotz2011,lopez-sanjuan2011,lopez-sanjuan2013}, staying high at least to z$\sim 3-4$ \citep[e.g.,][]{conselice2008,tasca2014}, and one would therefore expect that a fraction of the clumps is related to ex-situ major merging. 

When deriving volume average quantities like $f_{clumpy}$ it is critical to observe samples representative of the general star-forming population, with well defined physical parameters including luminosity, stellar masses and sizes, and following a well defined selection function. The size of irregular high redshift galaxies is very different when using their effective radius or a  total size \citep{ribeiro2016}, and searching for clumps using one or the other of these sizes to define a search area may lead to significantly different results on $f_{clumpy}$.  We show that total  sizes are larger by a factor of $\sim1.5-2$ than what previous findings  using $r_e$ claim, and that this difference is larger at higher redshifts \citep[comparing $z\sim2$ to $z\sim4$,][]{ribeiro2016}. Moreover, surface brightness dimming and luminosity evolution with redshift complicate the issue of the definition and identification of a clump. One must be aware of the impact of specific definitions and sample selection, for subsequent analysis to have a statistically robust meaning in a representative volume of the Universe at a given redshift.

In this paper we have used a sample of 1242 galaxies with secure spectroscopic redshifts from the VIMOS Ultra-Deep Survey (VUDS) and HST imaging in the COSMOS and ECDFS fields to measure $f_{clumpy}$. We used a clump definition expanded from \citet{guo2015}, counting all clumps within the isophote defining the total size $r_{T,100}$ and in a 20 kpc physical radius. In our clump definition we made no attempt at distinguishing between galaxy nucleus and a star-forming clump. The reasoning behind this choice of classification is that at these redshifts the nucleus of one galaxy may well be in itself a star-forming clump, as the timescales involved in clump formation and respective lifetimes can be comparable to the age of the galaxy we are observing. We then used the observed properties of the detected star-forming regions to infer their possible nature and formation mechanisms.

This paper is organized as follows. The sample and the data we used are described in Sect. \ref{sec:data_sample}. We present our method for identifying and characterizing clumps in Sect. \ref{sec:clumps_definition}. The evolution of the fraction of clumps is presented in Sect. \ref{sec:clumpyfrac}. We analyze the statistical properties of clumps in star-forming galaxies in Sect. \ref{sec:clump_physprops} where we explore the distribution of clumps in galaxies, their areas and luminosities and their spatial distribution.  The results are discussed in Sect. \ref{sec:discussion} and summarized in Sect.\ref{sec:summary}.
We use a cosmology with $H_0=70\mathrm{~km~s^{-1}~Mpc^{-1}}$, 
$\Omega_{0,\Lambda}=0.7$ and $\Omega_{0,m}=0.3$. 
All magnitudes are given in the AB system.


\section{Data and sample selection}\label{sec:data_sample}

{

We aimed to work with a sample of galaxies with accurate spectroscopic redshifts. This avoids the uncertainties generally related to photometric redshifts with typical accuracy at $z\sim3.5$ of $\delta z \sim 0.2-0.3$ when deriving important physical quantities like sizes, the luminosity of an isophote, or total luminosities and stellar masses. 

We then imposed that the galaxies with spectroscopic redshifts have deep HST imaging available. As algorithms for clump detection are sensitive to the redshift of the galaxy  (see Sect. \ref{sec:clumps_definition}), the HST images combined with the spectroscopic redshift of a galaxy enable the computation of reliable isophotal areas within which we identified clumps and quantify their properties. We defined as a clump any component  in a galaxy image that counts more than five connected pixels above a reference isophote (see Sect. \ref{sec:clumps_definition}).

We further restricted our sample above a given stellar mass defined as a function of redshift from the evolving mass function. This ensures that we followed a similar population of galaxies over the explored redshift range. This selection of the sample and associated data is described in detail in Sect. \ref{sect:vuds} to \ref{sect:mass}.

\subsection{Starting sample selection: Galaxies with spectroscopic redshifts from the VIMOS Ultra Deep Survey (VUDS)}
\label{sect:vuds}

We use the VUDS sample, the spectroscopic survey conducted by our team with $\sim10\,000$ objects targeted in an area of 1deg$^{2}$ on three separate fields: COSMOS, ECDFS and VVDS-02h \citep{lefevre2015}. The VUDS sample covers the redshift range $2<z<6+$, and produces a magnitude-selected sample of galaxies with $22.5\leq i_{AB} \leq 25$ and spectroscopic redshifts.

The spectra were obtained using the VIMOS spectrograph on the ESO-VLT \citep{LeFevre:03} covering, with two low resolution grisms (R=230), a wavelength range of $3650\AA < \lambda < 9350\AA$. The total integration time is  $T_{exp} \sim 14$h per pointing and grism.
Data processing was performed within the VIPGI environment \citep{scoddeggio2005}, and followed by extensive spectroscopic redshift measurements campaigns using the EZ redshift measurement engine  \citep{garilli2010}. At the end of this process each galaxy has a spectroscopic redshift measurement, and an associated redshift reliability flag. 

 In the following we analyze exclusively VUDS galaxies with the most reliable spectroscopic redshifts, that is, galaxies with spectroscopic redshift flags 2 and 9 ($\sim$75\% probability to be correct), and flags 3 and 4 (95 to 100\% probability to be correct). For more information on this sample we refer the reader to \citet{lefevre2015}.


  \subsection{Sample with imaging data in the VUDS-COSMOS and VUDS-ECDFS fields}\label{sec:imaging_data}

To reliably identify clumps in galaxies we use those galaxies in VUDS as defined in Sect. \ref{sect:vuds} which have deep high resolution images available from HST surveys. The COSMOS survey \citep{scoville2007,koekemoer2007}  has $\approx2\mathrm{deg^2}$ of HST ACS F814W (I band) coverage down to a limiting magnitude of 27.2 ($5\sigma$ point-source detection limit), and covers the entirety of the VUDS COSMOS area of $\sim1800$ square arcminutes \citep{lefevre2015}. This field provides the largest HST imaging data set of galaxies with spectroscopic redshifts. In the ECDFS we use the F814W images overlapping with the VUDS footprint over 675 square arcminutes \citep[see ][]{lefevre2015}. The F814W images cover the UV rest-frame from $\sim$2700\AA ~at z$\sim$2 to $\sim$1500\AA ~at z$\sim$4.5.

To probe the rest-frame optical morphology we use the smaller sample of galaxies in VUDS covered by the CANDELS HST imaging survey as presented in \cite{tasca2016}, a subset of the full CANDELS area \citep{grogin2011,koekemoer2011}. Even if small, the area  of $\sim270$ square arcminutes is important because the near-infrared coverage (with the F125W and F160W bands) allows us to compare optical rest-frame properties to UV rest-frame derived from the F814W imaging. Data in CANDELS reaches depths of $5\sigma$ of 28.4, 27.0 and 26.9 at F814W, F125W and F160W, respectively. 
The typical spatial resolution of these images ranges from $0.09$\arcsec ($0.8, 0.6$ kpc, at $z=2.0, 4.5$) for the F814W band up to $0.18$\arcsec ($1.5, 1.2$ kpc, at $z=2.0, 4.5$) in the F160W band.

\subsection{Selecting a stellar-mass limited sub-sample}
\label{sect:mass}

        \begin{figure*}
   \centering
   \includegraphics[width=\linewidth]{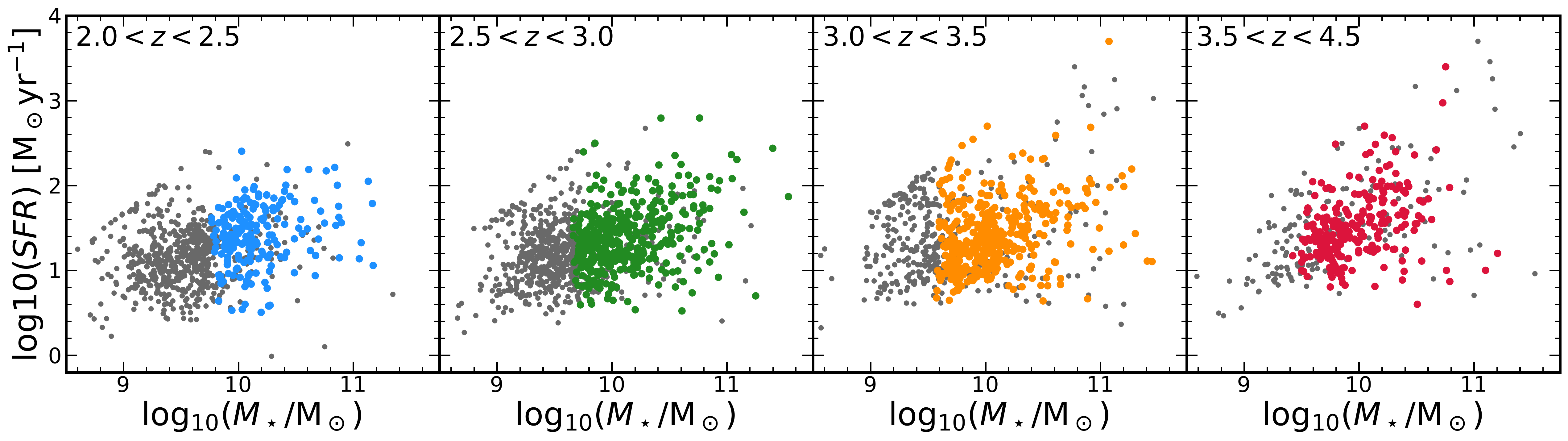}
      \caption{Stellar masses and star formation rates for the galaxies in the selected sample. In each panel, the light gray points show all VUDS galaxies with measured redshift in each bin. The colored circles show the 1242 galaxies in the selected stellar mass range and with correct redshift probability of $>75\%$ (see text for more details).}
         \label{fig:mass_sfr_plot}
   \end{figure*}

\begin{table*}
\centering
\begin{tabular}{|c|c|c|}
\hline
Property & COSMOS & ECDFS \\
\hline
 \multirow{2}{*}{Photometry} &  NUV,FUV,u,g,r,i,z,Y,J,H,K & U,B,V,R,I,z,J,H,K\\  
 & IRAC (3.6$\mu$m, 4.5$\mu$m)  & IRAC (3.6$\mu$m, 4.5$\mu$m) \\
\hline
Ages & \multicolumn{2}{|c|}{51 ages (in the range 0.1 Gyr to 14.5 Gyr)} \\
\hline
IMF & \multicolumn{2}{|c|}{\citet{chabrier2003}} \\
\hline
Metallicities & \multicolumn{2}{|c|}{Z=1, 0.4, 0.2 Z$_\odot$} \\
\hline
Dust extinction law &  \multicolumn{2}{|c|}{\citet{calzetti2000}, SMC-like law \citep{arnouts2013}} \\
\hline
$E(B-V)$  & \multicolumn{2}{|c|}{0, 0.05, 0.1, 0.15, 0.2, 0.25, 0.3, 0.4, 0.5, 0.6, 0.7}\\
\hline
Star formation histories & \multicolumn{2}{|c|}{Exponential ($\tau = 0.1,0.3,1,3,5,10,30$ Gyr) and delayed exponential (1,3 Gyr delay)}\\
\hline

\end{tabular}
\caption{ 
Details of the parameters used for our SED fitting. We use the publicly available photometry in the COSMOS field: GALEX - \cite{zamojski2007}; Subaru - \cite{capak2007}; Ultra-Vista - \cite{mccracken2012}; Spitzer-IRAC - \cite{sanders2007}. For the ECDFS we use the photometry compilation by \cite{cardamone2010}.
}
\label{tab:SED}
\end{table*}

In the final step of our sample selection we define a mass-selected sample from the sample defined in Sect. \ref{sec:imaging_data}, based on stellar masses computed from spectral energy distribution (SED) fitting on the large set of photometric data available in the COSMOS and ECDFS fields.
The SED fitting procedure closely follows the method described in \citet{ilbert2013} using the latest version of the Le Phare code (maintained by Olivier Ilbert), with the specifics for VUDS  detailed in \citet{tasca2015} and summarized in Table \ref{tab:SED}. The main parameter of interest in this paper is the stellar mass, M$_{\star}$ for which the median values of the probability distribution function are used. We refer the reader to \citet{ilbert2013} and \citet{thomas2016} for typical uncertainties on these quantities \citep[see also][]{tasca2015}. The statistical uncertainty on stellar masses is typically $\sim0.1$ dex. Systematic uncertainties due, for example, to the choice of initial mass function (IMF) can be as high as a factor of $\sim1.7$, if a \citet{salpeter1955} IMF is used instead of the \citet{chabrier2003} IMF.
The stellar mass selection (discussed below), color correction detailed in Sect. \ref{sec:clumps_isophote} and the k-correction used in Sect. \ref{sec:clumps_physic} rely on the parameters derived from the SED fitting with Le Phare. Using stellar masses as derived from another SED fitting code as for example, the GOSSIP+ code \citep[see ][]{thomas2016}, performing SED fitting on the joint photometry and spectroscopy data, has no impact on the results presented in this paper.

We present the stellar mass - star forming rate (SFR) relation of our sample divided in four redshift bins in Fig. \ref{fig:mass_sfr_plot}. It shows that our sample is representative of the high stellar mass population in VUDS and that it probes galaxies with typical median stellar masses of $10^{10}\mathrm{M_\odot}$ and ranging from $\log_{10}(M_\star/M_\odot)\gtrsim 9.35$ and up to $\log_{10}(M_\star/M_\odot)\gtrsim 11$. In terms of SFR, our galaxies are in the range $0.5\lesssim \log_{10}(SFR)\lesssim 3.0$ at all redshifts with median values of $\log_{10}(SFR)\sim 1.4$.

To follow the evolution of galaxy properties in a similar population as a function of redshift, broadly representing the same coeval population, we defined our sample imposing an evolving lower stellar mass limit set to follow the general stellar mass growth of star-forming galaxies. We defined the lower stellar mass limit in our sample as  the limit below which the VUDS sample is becoming incomplete at $z=4.5$, setting $\log_{10}(M_\star/\mathrm{M}_\odot)>9.35$. We then used the stellar mass function evolution from \citet{ilbert2013} together with the typical sSFR of VUDS galaxies \citep{tasca2015} to follow the typical stellar mass growth of VUDS galaxies, and defined the stellar mass selection threshold at different redshifts using
\begin{equation}
\log_{10}({M_\star/M_\odot}) > -0.204\times(z-4.5)+9.35.
\end{equation}
We refer to \citet{ribeiro2016} for more details on the choice of this mass selection in the context of the VUDS survey. In the higher redshift interval $z>4.5$ we selected all galaxies with spectroscopic redshift and stellar masses  $\log_{10}({M_\star/M_\odot}) > 9$ to build a sufficiently large sample to obtain a measurement of the clumpy fraction and clump properties, even if this sample is slightly incomplete.

\subsection{Final sample}

We identified a total of 1242 galaxies  with $2<z<4.5$  that satisfy the mass-selection and the redshift quality criteria in the VUDS survey with ACS F814W images, of which 1087 are in the COSMOS field and 155 are in the CANDELS-ECDFS field. We also examined a higher redshift sample of 96 galaxies with $4.5<z<6$, even if the available imaging data for this higher redshift sample is not deep enough to search for clumpy galaxies at the same luminosity limit than the lower redshift sample, and we take this into consideration in our analysis.

}

\section{Defining clumps in galaxies}\label{sec:clumps_definition}

\begin{figure*}
\centering
\includegraphics[width=0.97\linewidth]{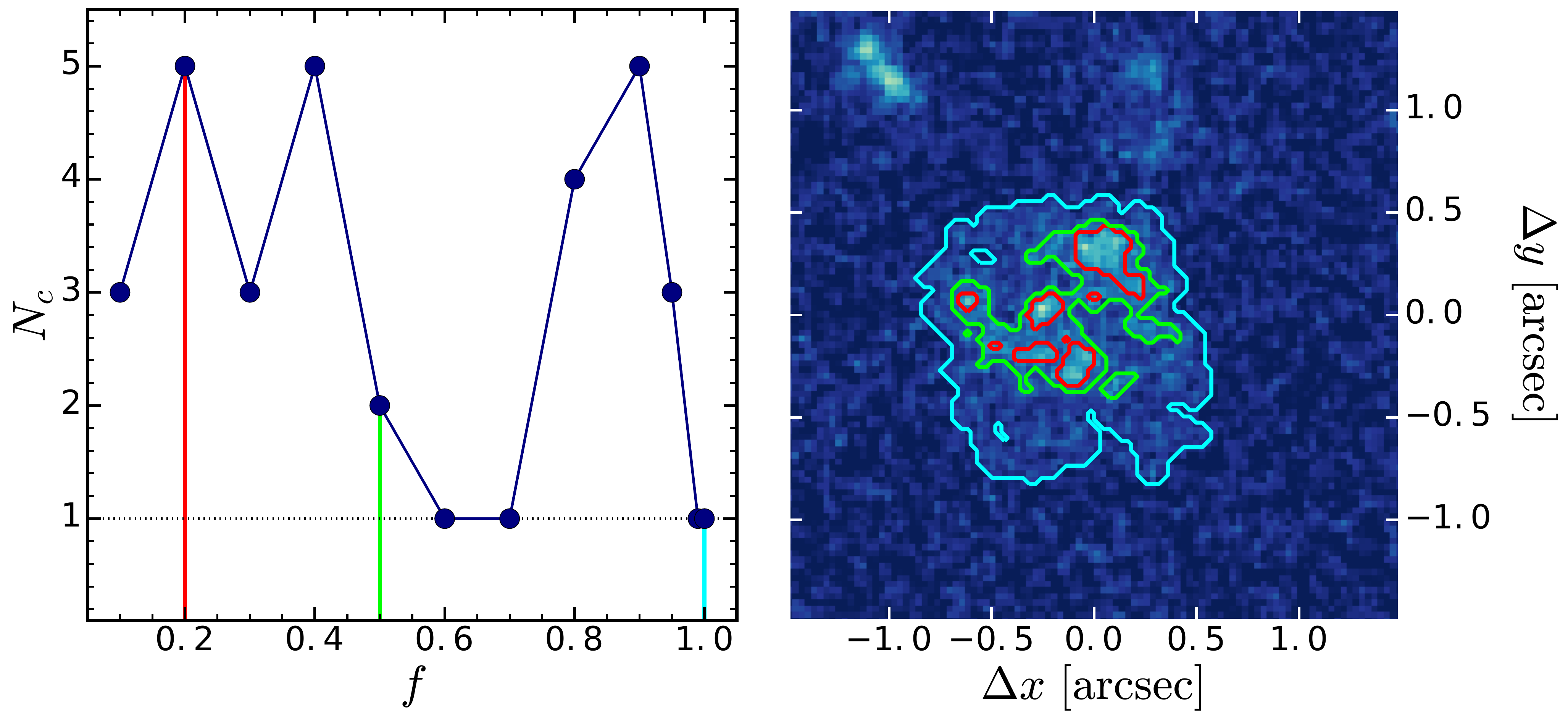}
\caption[Methodology for detecting clumps]{Exemplification of the method for detecting galaxy clumps. On the left we show the number of disconnected regions with more than ten pixels as a function of flux level cut. When $f=1$ the isophote limit represents the limiting area of the galaxy. On the right we show the contours corresponding to three different flux level cuts at 20\%, 50\% and 100\% of total flux that are identified in the same color as the vertical lines in the left panel. }
\label{fig:clumpy_method}
\end{figure*}

There are various methods in the literature that have the goal of detecting and characterizing galactic clumps at high redshift  ranging from visual classification to automated algorithms \citep[e.g.,][]{elmegreen2007,puech2010,guo2012,wuyts2012,tadaki2014,guo2015,shibuya2016}. The clump detection algorithms generally involve the subtraction of a smoothed version of an image from the original and then search for positive detections in the residual image that can be considered as galactic clumps, a simple and robust method. In the work by \citet{guo2015} and \citet{shibuya2016} a clump is detected if five connected pixels are detected at $2\sigma$ above the observed image background.  A surface brightness limit defined with respect to the background corresponds to a fainter intrinsic (absolute) luminosity at lower redshift than at higher redshift because of surface brightness dimming varying as $(1+z)^{-3}$ \citep{ribeiro2016}. For instance, between z=2 and  z=4 using the same surface brightness limit defined with respect to the background leads to almost a factor of five difference in the intrinsic surface brightness of these galaxies. Counting clumps may therefore be significantly affected depending on the definition of the limiting surface brightness limit. The overall lowering of the clump fraction with increasing redshift as reported by previous studies may simply be due to the bias produced by using the same observed surface brightness at all redshifts when identifying clumps. Because of the limits in detection at the highest redshifts one probes only the brighter and rarer clumps, while at the same limiting surface brightness one finds fainter clumps which are more numerous, hence mimicking a clump fraction decrease with increasing redshift. 

To follow on from these considerations, we followed a slightly different approach in this paper than in the recent studies of \citet{guo2015} and \citet{shibuya2016}. We imposed a surface brightness limit when detecting clumps  which has the same intrinsic luminosity for all measured galaxies, whatever their redshift. We have therefore taken into account surface brightness dimming differences, which are important over the redshift range considered. Furthermore, the average luminosity of a galaxy population changes with redshift as shown by luminosity function studies (and the evolution of the luminosity density), and we further adjusted the limiting surface brightness when searching for clumps taking into account this redshift dependent luminosity evolution.  

In this section we describe the method we follow to identify clumps and measure their luminosity and size. We also reproduced the method used by \citet{guo2015} and \citet{shibuya2016} for comparison with our results on the same galaxy sample.

\subsection{A clump finding method consistent at all redshifts}

We summarize our method in the logical steps below.

\begin{enumerate}
 \item We smoothed a $6\arcsec \times6\arcsec$ image centered on the VUDS galaxy with a Gaussian kernel of $\sigma=1$ pixel (0.03\arcsec);
 \item We computed the surface brightness threshold corresponding to the galaxy, from a reference {\it intrinsic} threshold imposed on all galaxies, that takes into account the relative surface brightness diming as well as the average luminosity evolution of the galaxy population;
\item We created a segmentation map that encodes which pixels are found to be above the defined surface brightness threshold;
 \item From the segmentation map we selected the area of connected pixels above the adopted surface brightness threshold which contains the brightest pixel within a 0.5 arc-second radius from the target coordinates;
\item Using the selected pixels, we computed the total flux of the galaxy by summing up the flux in all the pixels;
\item  We iterated (using 1000 steps) a flux level $f_i$ from the maximum to the minimum flux detected. At each step we computed the number of disconnected regions (defined from all pixels with fluxes greater than $f_i$) within the segmentation map  and stored the position of the local maximum associated with each disconnected region (see Fig. \ref{fig:clumpy_method}).
\item We defined the number of clumps as the number of disconnected regions detected above the isophote level corresponding to $80\%$ of the total galaxy flux, which are associated with the different local maxima.
\end{enumerate}
We repeated this process for all galaxies found in the image stamp. We note that we have only considered clumps with a detection area greater that ten pixels. This minimum area for clump detection ensures that we have excluded the contribution from random fluctuations of photon counts and also imposes a lower limit of $\sim 3$ pixels (roughly the FWHM of the PSF) on the separation of any given two clumps that we can detect.

We describe some of the key elements of this algorithm in Sect. \ref{sec:clumps_isophote} to \ref{sec:kcorr}.

\subsection{Limiting surface brightness: isophotal threshold}\label{sec:clumps_isophote}

One of the key elements of this algorithm is the limiting isophote at which we define the total extent of a single galaxy. For our method, it is defined as
\begin{equation}
k = k_p \left( \frac{1+z}{1+z_p} \right)^{-3} \times 10^{-0.4(I_\mathrm{obs}-B_\mathrm{rest})} \times \frac{L(z)}{L(z_p)},
\label{eq:thresh}
\end{equation}
where $k_p$ (pivot threshold) is the value of $k$ at $z=z_p$ (pivot redshift, with $z_p=2$ in this work) and $I_\mathrm{obs}-B_\mathrm{rest}=0$, $z$ is the spectroscopic redshift, $I_\mathrm{obs}$ is the apparent magnitude of the galaxy,  $B_\mathrm{rest}$ is the apparent rest frame magnitude of the band $B$, and $L(z)$ is the typical luminosity at redshift $z$ (see Eq. \ref{eq:lumevo} for more details). Equation \ref{eq:thresh} is composed of three different terms:

{\bf The dimming correction term}, $\left[ (1+z)/(1+z_p) \right]^{-3}$, which accounts for the fact that due to the expansion of the Universe the observed number of photons of a source with constant intrinsic luminosity scales with $(1+z)^{-3}$ \citep[e.g.,][]{ribeiro2016}.

{\bf The color correction term}, $10^{-0.4(I_\mathrm{obs}-B_\mathrm{rest})}$,  which accounts for the fact that we are not  observing galaxies exactly at the same rest-frame wavelengths and corrects for the relative brightness based on the slope of the SED close to the band we are observing.  To be consistent in our large redshift range we chose to use rest-frame Far Ultraviolet (FUV) magnitudes to compute the color correction term since it is the central wavelength at the peak of the redshift distribution of our sample.

{\bf The luminosity correction term}, ${L(z)}/{L(z_p)}$,  which accounts for the average luminosity evolution in galaxies. 
$L(z)$ is derived from the $L^*$ values computed in FUV by \cite{reddy2009} at $z<3$ and by \cite{bouwens2015} at $z>4$. We defined a luminosity evolution which is steeper below $z<3$ and flattens at $z>3$:
\begin{equation}
L(z)=
\begin{cases} 
       10^{-0.4(-0.36z)} & z\leq 3 \\
      2.25\times 10^{-0.4(-0.07z)} &z>3 \\
   \end{cases}
.
\label{eq:lumevo}
 \end{equation}

 In Eq. \ref{eq:thresh} there are two parameters that are correlated. The value $k_p$ represents the detection threshold, in units of $\sigma$, at $z=z_p$. In this study we set $z_p=2$ and used several values of $k_p$ throughout the paper. In particular we used $k_p=1.5,2.0$ for $2<z<4.5$ and $k_p=5.0$ for $4.5<z<6$. Since the value of $k$ can vary by a factor of $((1+2)/(1+6))^{-3}\sim0.08$ from $z=2$ to $z=6$ based only on surface brightness dimming we chose different values of $k_p$ to allow for an estimate of the total extent of galaxies at the lower limit while preventing noise contaminated detections at the high redshift end.

\subsection{Luminosity and sizes}\label{sec:clumps_physic}

In order to obtain further information on the physical characteristics of clumps we adapted the method defined by \citet{freeman2013} to compute the intensity parameter {\it I} giving the luminosity of each clump.
This algorithm defines groups of pixels associated with a local maximum inside a segmentation map. First, the image was smoothed via a Gaussian kernel with a given $\sigma$ (we use $\sigma=1$pixel). Then, for each pixel in the segmentation map, we computed the corresponding local maximum via maximum gradient paths. This means that for each pixel we computed the intensities of the eight surrounding pixels and then moved to the pixel with maximum intensity. We continued this process until a local maximum was reached and none of the eight neighbor pixels had higher intensities than the one we were considering. So, a pixel group is defined as the collection of pixels linked to a particular local maximum.

Once we had all the pixel groups defined, we matched the local maximum of each group to the local maximum of the detected clumps as described in Sect. \ref{sec:clumps_definition}. Then we computed the luminosity and area associated with each clump by counting all the light and pixels in each group, respectively. It is important to stress that since we stopped at the border of two contiguous groups of pixels, the derived areas and luminosities are lower limits to the real values. This happens since we only considered flux from pixels where the contribution from the clump being considered is dominant with respect to neighboring clumps.

With the knowledge of the spectroscopic redshift from VUDS, we were then able compute physical areas in kpc$^2$. Absolute magnitudes are computed using the flux measured from the F814W images. We corrected the magnitudes to FUV rest-frame using the k-correction derived for the galaxy from SED fitting. This means that the clump magnitude is derived as
\begin{equation}
M_\mathrm{clump,FUV} = M_\mathrm{clump,F814W} + k_\mathrm{corr,FUV}.
\end{equation}

\subsection{Resulting clumps and pair identification}\label{sec:result_clump}

At the end of this process a list of galaxies was identified within a  20 kpc radius, a value commonly used as two equal masses bound by gravity  would likely merge in a dynamical timescale of $\sim$1 Gyr   \citep[][]{lopez-sanjuan2013,tasca2014}, and a list of clumps was identified in each galaxy.  Our algorithm may identify several galaxies within the search radius of 20 kpc, as pixels groupings isolated from other pixel groupings, that is, these groups of pixels are not connected to each other at the isophote considered. It is well possible that if deeper images were available, allowing us to lower the search isophote, some of these pixel groupings could be connected and possibly associated to a single galaxy. 

\begin{figure*}
\centering
\includegraphics[width=0.97\linewidth]{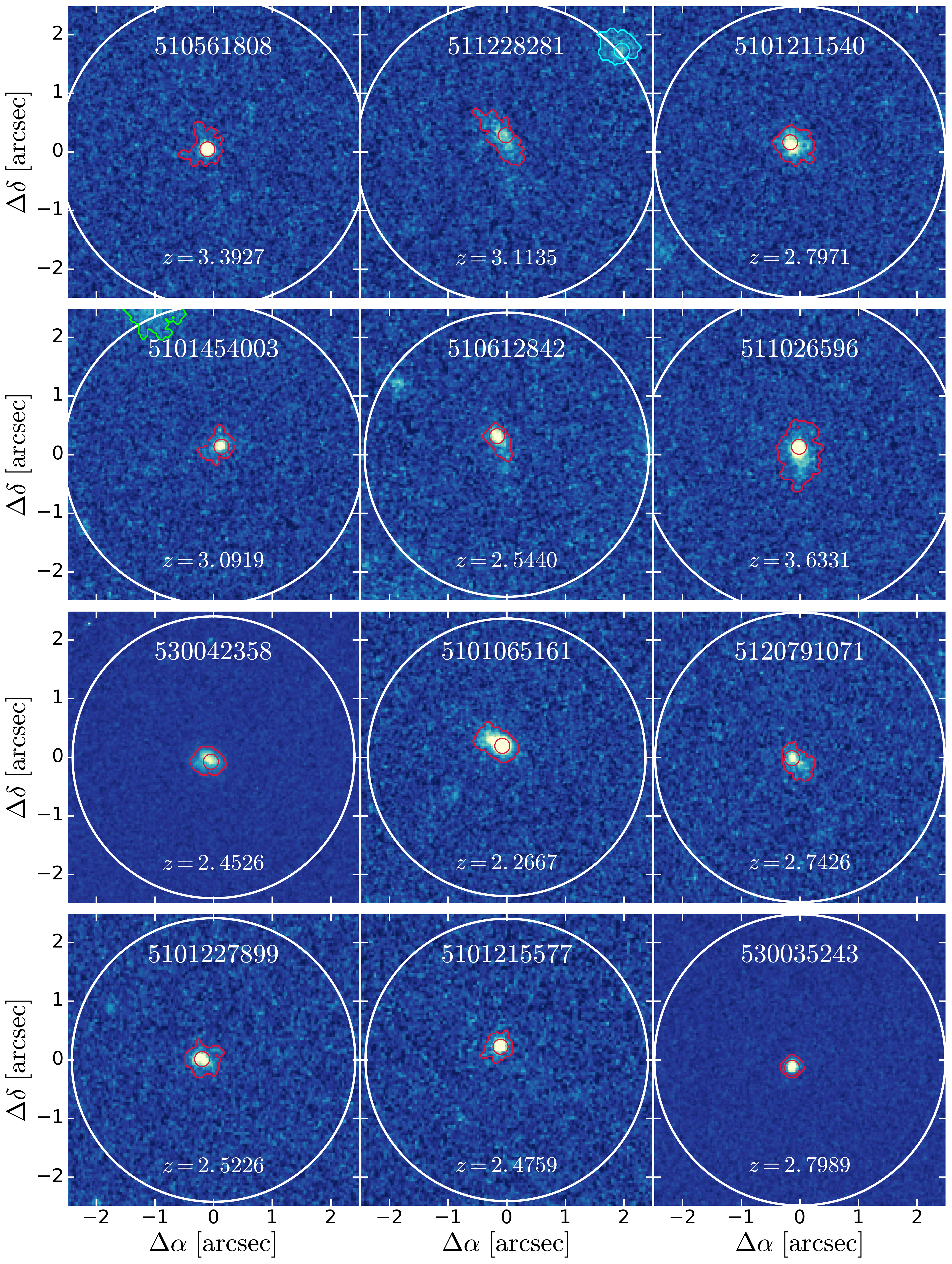}
\caption{Twelve examples of galaxies for which only one clump was detected. The white circle shows the 20 kpc projected search radius. The contours represent the detection isophote of each galaxy identified within this search radius. Each circle shows the position of a detected clump within the isophote of the same color. VUDS targets are identified with red contours and circles.}
\label{fig:clumpy_examples1}
\end{figure*}

\begin{figure*}
\centering
\includegraphics[width=0.97\linewidth]{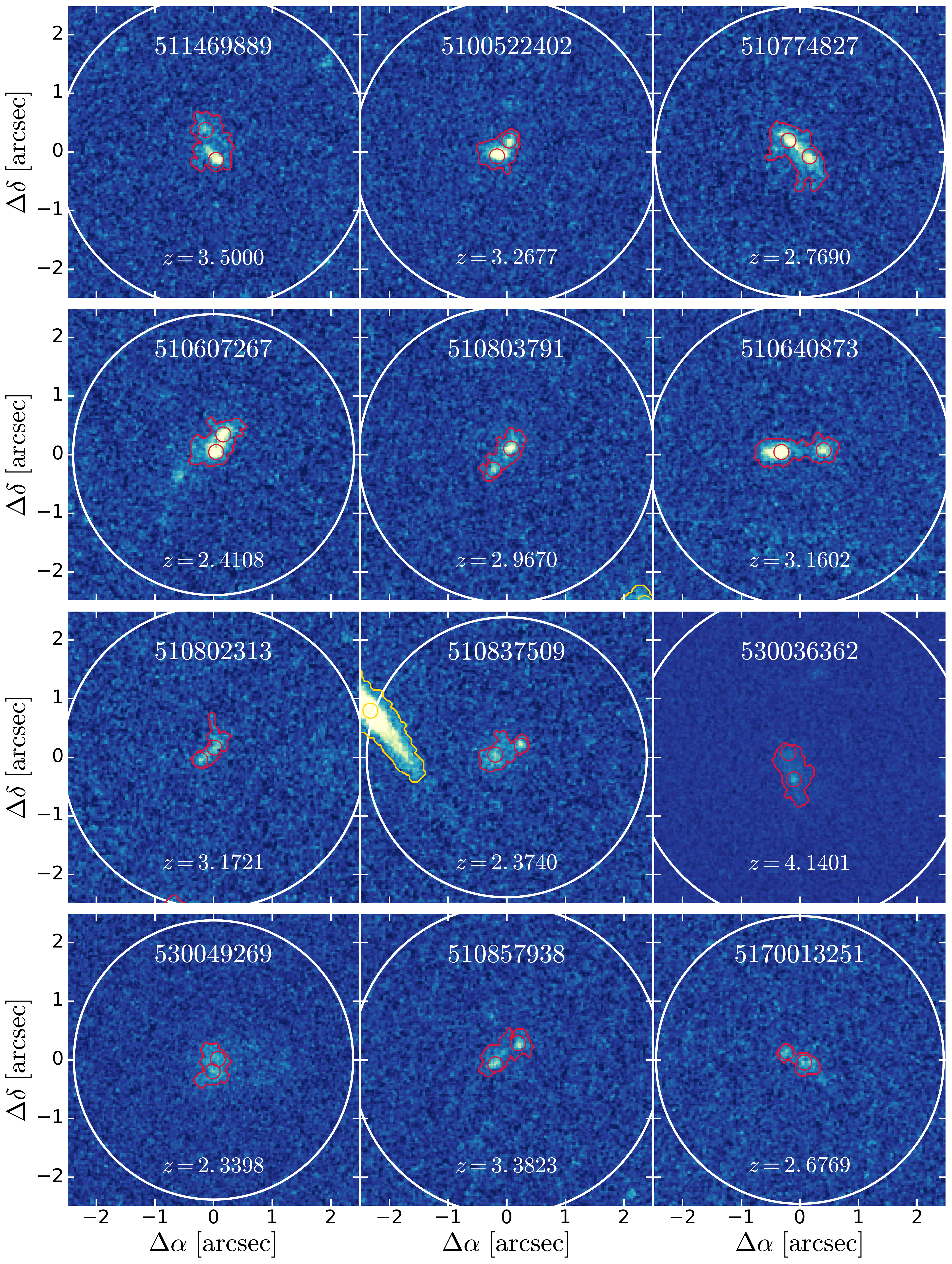}
\caption{Twelve examples of galaxies with two clumps detected. The top six examples show cases for which at least one  bright clump was detected whereas the bottom six examples show pairs of faint clumps. The white circle shows the 20 kpc projected search radius. The contours represent the detection isophote of each galaxy identified within this search radius. Each circle shows the position of a detected clump within the isophote of the same color. VUDS targets are identified with red contours and circles.}
\label{fig:clumpy_examples2}
\end{figure*}

\begin{figure*}
\centering
\includegraphics[width=0.97\linewidth]{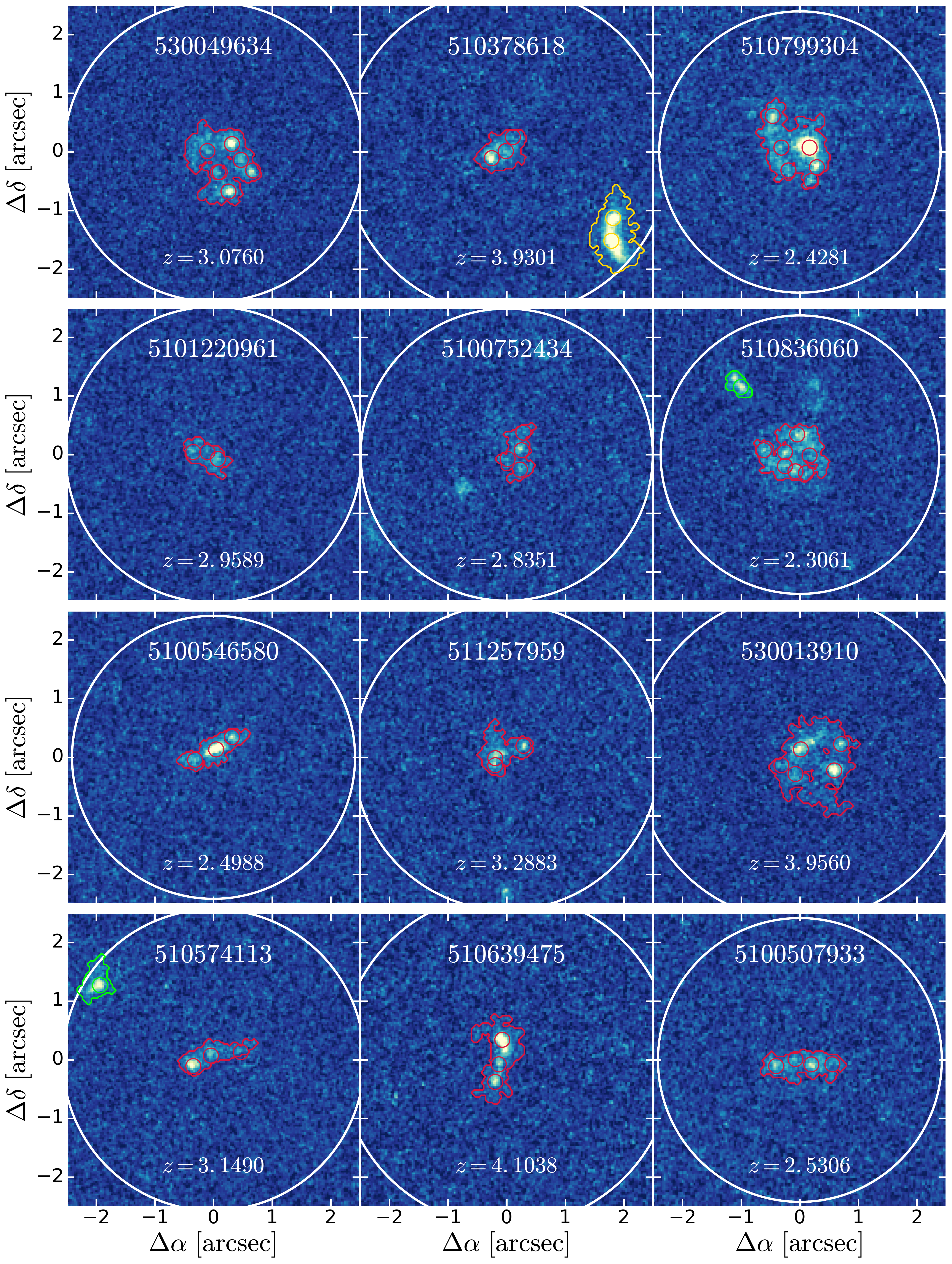}
\caption{Twelve examples of galaxies with more than two clumps detected. The white circle shows the 20 kpc projected search radius. The contours represent the detection isophote of each galaxy identified within this search radius. Each circle shows the position of a detected clump within the isophote of the same color. VUDS targets are identified with red contours and circles.}
\label{fig:clumpy_examplesM}
\end{figure*}

To avoid ambiguous interpretation, all our results on galactic clumps take into account only clumps that are associated with the galaxy that was the target of the VUDS spectroscopy.  We have spectroscopic information (important to compute $k$ in Eq. \ref{eq:thresh}) only for these sources  as well as physical parameters from SED fitting only for these galaxies.  However we keep information about other galaxies detected within 20 kpc to compute a pair fraction at the same redshifts, as discussed in Sect. \ref{sec:discussion}. Several examples of  clumps detected in VUDS galaxies are shown in Figs. \ref{fig:clumpy_examples1} - \ref{fig:clumpy_examplesM}.

To transform our pair fraction into a merger fraction we corrected the observed pair count for line of sight contamination using a simple statistical argument. Starting with the expected number of galaxies within a square degree, as derived by \citet{capak2007}, we computed the probability of a galaxy of magnitude $m$ to be found at a distance $r$ of the VUDS target. The total number of galaxies brighter than magnitude $m$ in one square degree was computed  by integrating the number counts, $N(<m)$ computed by \citet{capak2007}. We then estimated how many galaxies we expected to find within $\pi\times r^2$. This is summarized by
\begin{equation}
p  =  \frac{\pi r^2}{3600\times3600} \times \int^m_{15} N(<m^{\prime}) dm^{\prime}.
\end{equation}
To correct for line of sight contamination, each galaxy of magnitude $m$ is weighted by its probability of being a contaminant, $0<p<1$, when we computed the pair fraction.

\subsection{False clump detections}

\begin{figure}
\centering
\includegraphics[width=\linewidth]{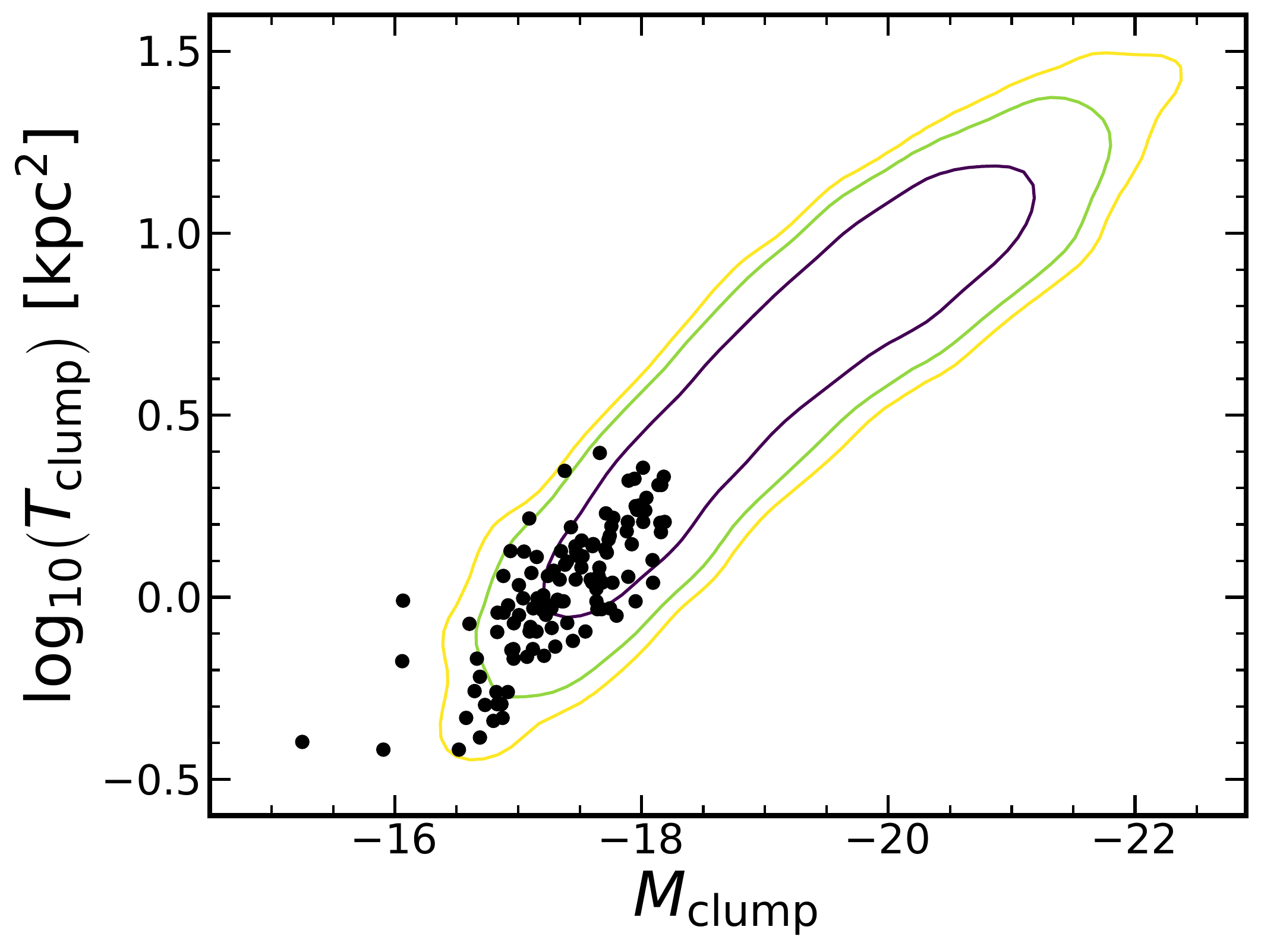}
\caption{Black points show the distribution in areas and magnitudes for the negative clumps detected applying the same method on the inverted galaxy image. The contours depict the 68\%, 90\%, and 95\% levels of the positive clump detections in our data. Most of the negative detections are found at $M_\mathrm{clump}>-18$ which is fainter than the peak for the fainter clumps that is observed in the magnitude distribution (see Fig. \ref{fig:clumps_physics_2}). This indicates that the smaller clumps with $M_\mathrm{clump}<-18$ are mostly real physical clumps.}
\label{fig:clumps_physics_error}
\end{figure}

To estimate the impact of false detections on the inferred distributions of areas and luminosities we performed a simple test. We ran our clump detection algorithm using the same threshold as before on the negative image of each galaxy and computed the area and magnitude of each negative clump which is, by its nature, a false peak detection. We can see in Fig. \ref{fig:clumps_physics_error} that the inferred distribution of false clumps is found at small areas (peaking at $\sim$1 kpc$^2$) and faint luminosities (peaking at $M_\mathrm{F814W}\sim -17.4$ mag) as expected. 

Most of the negative detections are found at $M_\mathrm{clump}>-18$ which is fainter than the peak for the fainter clumps that is observed in the magnitude distribution (see Fig. \ref{fig:clumps_physics_2}). This indicates that the smaller clumps with $M_\mathrm{clump}<-18$ are mostly real physical clumps.

\subsection{Impact of the spatial resolution and morphological k-correction}
\label{sec:kcorr}

We assumed that clumps are evenly detected across the rest-frame wavelength range that we probed at different redshifts. We tested this assumption by computing the same quantities on a version of the F814W images degraded to the same spatial resolution as the H-band CANDELS images,  and on a sub-sample of galaxies for which H-band imaging from CANDELS is available. To degrade the I-band imaging data we re-scaled the image to match the pixel-scale of the H band and then convolved the image with a Gaussian kernel to match the PSF resolution of that image.

We find that the clumpy fraction in the degraded I-band images is found to be approximately three times smaller  than what we find for the non degraded images. In the H-band imaging we find that the clumpy fractions are fully consistent with the values we get for the degraded I-band imaging data, possibly slightly lower. We therefore argue that the differences we find when comparing I- to H-band data can be explained mostly by the difference in spatial resolution which, as expected, merges clumps that are too close to be resolved. We note however that we find no evidence for the existence of a morphological k-correction that needs to be applied to $f_{clumpy}$.

\subsection{Comparing with the \citet{guo2015} and \citet{shibuya2016} clump finding method}
\label{sec:guo_method}

For comparison purposes, we have also implemented the same methodology as used in recent studies \citep[e.g.,][]{guo2015,shibuya2016}.  As discussed at the start of this Sect. \ref{sec:clumps_definition} the method used by \citet{guo2015} and \citet{shibuya2016} is based on selecting clumps with respect to the local image background. We implement the following algorithm:
\begin{enumerate}
\item \textsc{SExtractor} is run on a $7\arcsec$ by $7\arcsec$ image to obtain the segmentation map.
\item The segmentation map is dilated using a binary operation that effectively smooths and enlarges by a maximum of three pixels the outer edges of the segmentation map.
\item  A small cutout of the original image is selected to adapt to the size of the segmentation map.
\item A boxcar filter (ten pixels width) is applied to the new cutout.
\item A residual image is produced by subtracting the filtered image to the original one.
\item The residual background flux is measured using a 3$\sigma$-clipping method.
\item All pixels that have fluxes smaller than 2$\sigma$ of the calculated background are masked.
\item We run a new detection algorithm on the masked residual image to detect clumps.
\end{enumerate}
All regions in the masked residual image that have more than five connected pixels are counted as single clumps. We also computed the magnitude of each clump using a three-pixel aperture.

To compute the final clump fraction with this method, we selected only galaxies for which GALFIT is converging into a reliable fit \citep[see][]{ribeiro2016} and with an effective radius $r_e>0.2\arcsec$ and axis ratio $q>0.5$. We selected galaxies with stellar masses in an intermediate range  ($9.8<\log_{10}(M_\star/M_\odot)<10.6$). As in  \citet{guo2015} we also included single off-center clumps when computing the clumpy fraction. Finally, we only considered clumps which contain more than 8\% of the galaxy light. We argue however that relying on GALFIT results for the final sample selection is limiting the potential for finding highly disturbed systems as well as very compact sources. It may also be that the structural parameters that we would rely on for that selection can be biased toward the brightest clump and as such not being representative of the entire galaxy, as we show in \citet{ribeiro2016}. Thus we believe that our methodology, which allows for a morphologically unbiased sample and includes a specific treatment of the apparent evolution caused by surface brightness dimming effects, can be complementary to that followed by \citet{guo2015} and \citet{shibuya2016} to provide a new perspective on the clumpy nature of high redshift galaxies. We compare the results obtained with this method to our results in Sect. \ref{sec:results_guo}.

\section{The evolution of the clumpy fraction with redshift}\label{sec:clumpyfrac}

\subsection{The distribution of the number of clumps per galaxy}\label{sec:distribution}

\begin{figure*}
\centering
\includegraphics[width=0.48\linewidth]{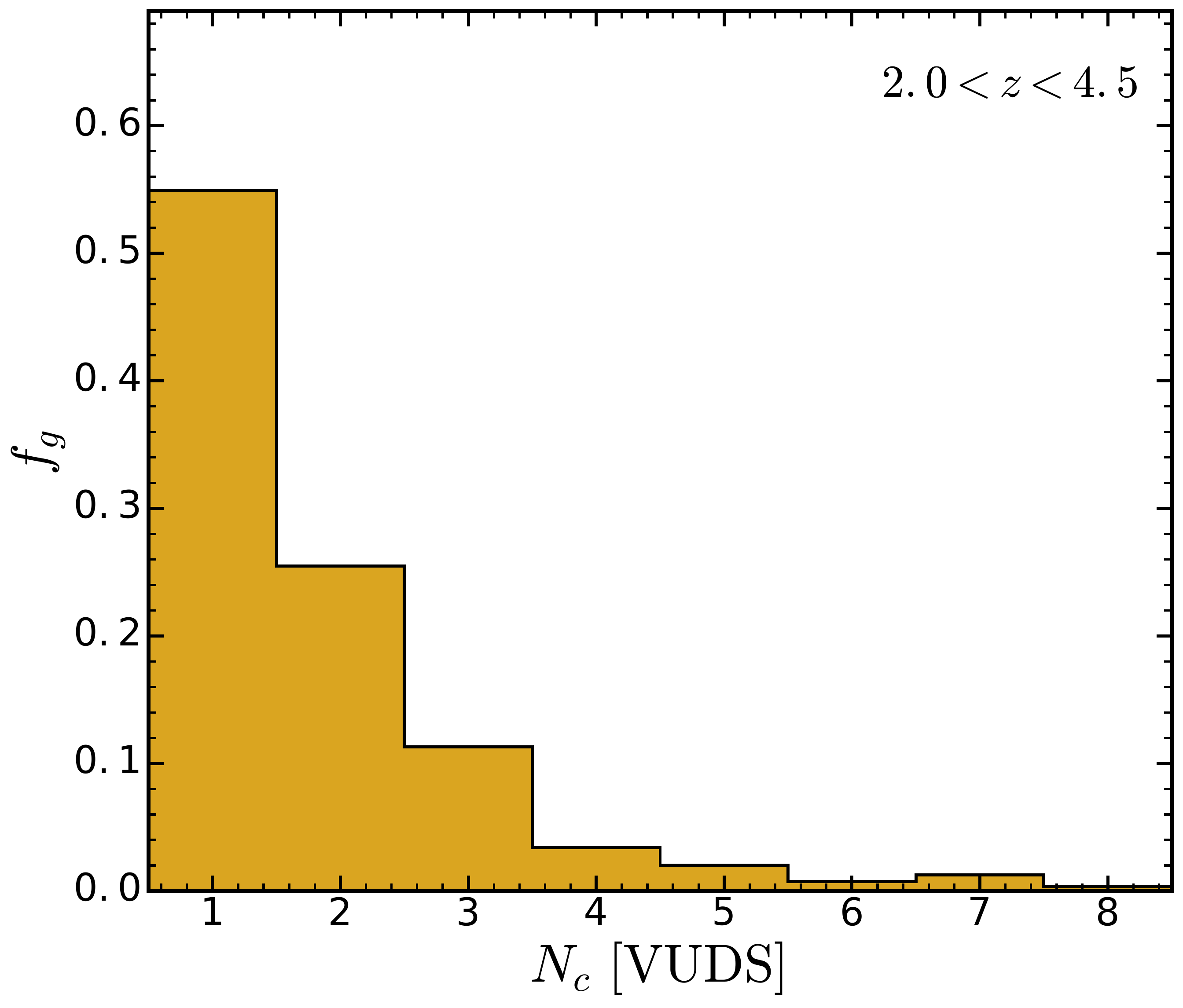}
\includegraphics[width=0.48\linewidth]{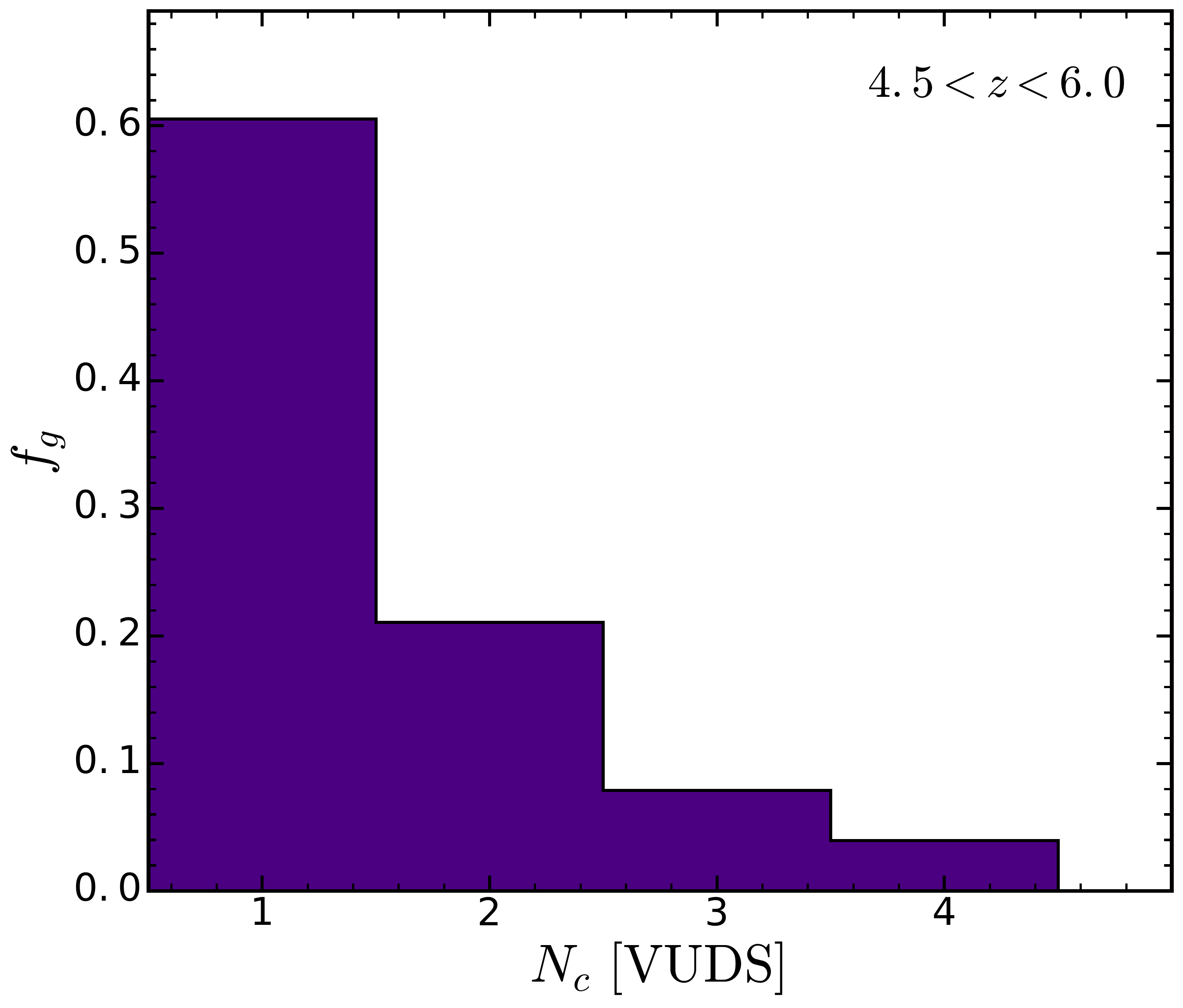}
\caption{Number of galaxies as a function of the number of clumps detected within the selected isophote. We note that we include in this plot single clump galaxies which correspond to galaxies with no substructure or a single central clump. We plot this quantity for two redshift intervals each with its respective value of $k_p=1.5$ for $2<z<4.5$ in gold (left panel) and $k_p=5$ for $4.5<z<6.0$ in purple (right panel). On each panel we show the number of clumps for the VUDS targets.}
\label{fig:clumps_in_vuds}
\end{figure*}

We compare in Fig. \ref{fig:clumps_in_vuds} the number of galaxies as a function of the number of detected clumps per galaxy in two  different redshift ranges. We consider only clumps that reside inside the isophote belonging to the VUDS target when showing this histogram. We find that the fraction of multi-clump systems decreases as the number of clumps increases.   Systems with only one clump (i.e., a single clump in a smooth galaxy at the observed resolution) are the dominant population ($\sim55\%$  at $2<z<4.5$ and $\sim61\%$ at $4.5<z<6.0$). We find that the fraction of galaxies with only two clumps dominates the distribution of clumpy galaxies with $\sim25\pm2.7\%$ double clumps galaxies at $2<z<4.5$ and $\sim21\pm7.7\ \%$ at $4.5<z<6.0$. Galaxies with three clumps are about $\sim11\pm1.6\%$ at $2<z<4.5$ and $\sim8\pm4.1\ \%$ at $4.5<z<6.0$. Galaxies with four or more clumps amount to only $\sim7\pm1.2\%$ at $2<z<4.5$ and $\sim9\pm4.5 \%$ at $4.5<z<6.0$. Several examples of galaxies with one, two, or more clumps may be found in Figs. \ref{fig:clumpy_examples1} - \ref{fig:clumpy_examplesM}, respectively.

\subsection{The fraction of clumpy galaxies}\label{sec:fraction}

\begin{table*}
\centering
\begin{tabular}{|c|ccccc|}
\hline
$k_p$ & $2.0<z<2.5$ & $2.5<z<3.0$ & $3.0<z<3.5$ & $3.5<z<4.5$ & $4.5<z<6.0$\\
\hline
1.5 &$0.34 \pm 0.08$ & $0.43 \pm 0.07$ & $0.46 \pm 0.07$ & $0.54 \pm 0.09$ & - \\
2.0 &$0.28 \pm 0.08$ & $0.32 \pm 0.06$ & $0.38 \pm 0.07$ & $0.47 \pm 0.08$ & - \\
5.0 &$0.08 \pm 0.05$ & $0.08 \pm 0.04$ & $0.12 \pm 0.04$ & $0.25 \pm 0.06$ & $0.39 \pm 0.12$\\
\hline
\end{tabular}
\caption[Clumpy fraction as a function of redshift]{Clumpy fraction as a function of redshift for three different values of $k_p$. In each entry we show the clumpy fraction for a given redshift bin and threshold normalization value $k_p$ and the respective Poisson derived errors.}
\label{tab:clumpy_frac}
\end{table*}

We defined any source that has more than one clump detected within its isophote as a clumpy galaxy. To compute this fraction as a function of redshift, we considered only clumpy (and non-clumpy) galaxies within VUDS for which we have reliable information on the spectroscopic redshift. This ensures that we did not smear our results between redshift bins by including galaxies with uncertain or incorrect redshifts. We find that the stellar mass distribution of clumpy and non-clumpy galaxies are indistinguishable from one another, so that there is no particular population bias in defining clumpy galaxies as those with more than one clump.
We stress that the clumpy fraction with our method depends on the selected isophote (value of $k_p$) and its dependency on a color correction, as allowed by the depth of available images. 

\begin{figure*}
\centering
\includegraphics[width=\linewidth]{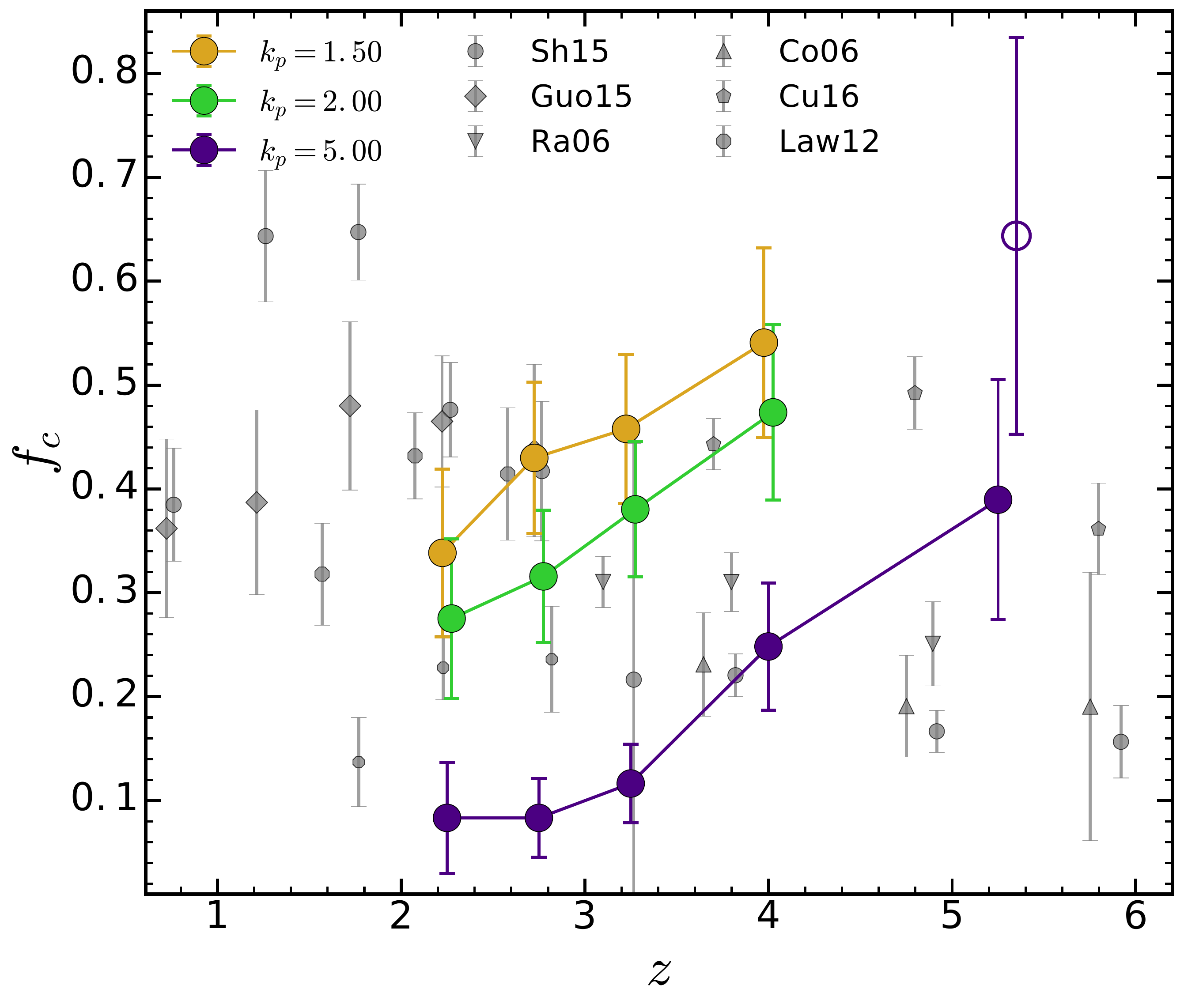}
\caption{Fraction of clumpy galaxies within the VUDS sample as a function of spectroscopic redshift. Different large colored circles (gold, green, and purple) represent different values of the normalization $k_p$. The smaller gray symbols represent values from the literature for clumpy galaxies: circles \citep{shibuya2015}; squares \citep{guo2015}; and for asymmetric galaxies: inverted triangles \citep{ravindranath2006}; triangles \citep{conselice2009}; octagons \citep{law2012}; pentagons \citep{curtis-lake2016}. For $k_p=5$, the isophote at $z<4$ is very high and as we only probe the brightest regions of galaxies we miss those clumps outside the search isophote, therefore lowering the fraction of clumpy galaxies. The empty purple circle at $z\sim5$ is the extrapolation of the clumpy fraction as if it was measured at $k_p=1.5$ (see Sect. \ref{sec:fc_highz} for details).}
\label{fig:clumpy_fraction}
\end{figure*}

We find that the clumpy fraction rises with redshift for all values of $k_p$ that we have tested as well as when using Near Ultraviolet (NUV) as the base rest-frame band. 
Due to the asymmetric nature of most galaxies at high redshift \citep[e.g.,][]{huertas-company2015,curtis-lake2016} and since deriving a center and a size can lead to different results when using different assumptions \citep[e.g.,][]{ribeiro2016}  we did not includie in our sample of clumpy galaxies those sources with an off-center clump as defined by, for example, \citet[][]{guo2015,shibuya2016}. The clumpy fraction evolution with redshift is shown in Fig. \ref{fig:clumpy_fraction} and summarized in Table \ref{tab:clumpy_frac} and is discussed below.

\subsubsection{Clumpy fraction in $2<z<4.5$}

At $2<z<4.5$ we observe that the clumpy fraction remains in the range $35$ to $55\%$ at all redshifts. This fraction is observed to be rising from low to high redshifts. We also report the clumpy fraction for two different values of $k_p$  showing that the absolute value does indeed depend on the isophote used.  Increasing the value of $k_p$ lowers the probability of detecting faint clumps and it is no surprise that the fraction of clumpy systems decreases as the value of $k_p$ increases. We find, regardless of the value of $k_p$ adopted, a factor $\ga2$ increase in the clumpy fraction from the lowest to the highest redshifts measured in our sample.
The rising evolution with redshift of the clumpy fraction observed using our method is different from results obtained in other studies \citep[e.g.,][]{guo2015,shibuya2016}. A possible explanation for these differences is that the lower fraction of clumpy galaxies that we observe at $z\sim 2-3$ is the result of the evolving isophote threshold defined in Sect. \ref{sec:clumps_definition}, including luminosity evolution. This imposes a higher threshold at lower redshifts than studies using a fixed apparent luminosity threshold, resulting in a reduced area in which to search for clumps and therefore a lower number of clumps at the lower redshift end. Our choice of taking into account luminosity evolution in defining the search isophote is motivated by the objective of counting clumps within similar physical areas while stellar populations brighten in galaxies when moving from the highest redshifts in our sample down to the peak in star formation rate density. When selecting a fixed apparent isophotal level like in \citet{guo2015} and \citet{shibuya2016}, the global brightening of galaxies enables to count clumps in larger galaxy areas making it more difficult to relate clump measurements at different redshifts.

\subsubsection{Clumpy fraction $4.5<z<6$}\label{sec:fc_highz}

At $z>4.5$ the depth of the data only allows the use of a higher value of $k_p$, as shown in Fig. \ref{fig:clumpy_fraction}. At these redshifts it is also necessary to use a brighter isophote limit for the clumps detection to avoid excessive contamination from noise. With these constraints we observe a clumpy fraction of $20-40\%$ in the redshift range $4.5<z<6$. 

It is clear from our results at lower redshifts that if the depth of the data was enabling us to use a fainter isophote we would likely observe a higher clumpy fraction in this redshift bin too. Since we have computed the clumpy fraction at all redshifts for all values of $k_p$ shown in Fig. \ref{fig:clumpy_fraction} we attempted to compute a correction to estimate the clumpy fraction in $4.5<z<6$ as if it was measured with $k_p=1.5$. To do that we fitted a linear curve to the  quantity $\frac{f_{c,K}}{f_{c,5}}(z)$
which quantifies the underestimation of the clumpy fraction measured at $k_p=5$ with respect to that measured at $k_p=K$. We then applied this correction factor measured at $z=5.25$ to correct the  clumpy fraction measured for $k_p=5.0$. In doing so, we estimated that the clumpy fraction at $4.5<z<6.0$ is $\sim64\pm19\%$ when corrected for $k_p=1.5$.

\subsubsection{Comparison of the clumpy fraction as derived from the \citet{guo2015} method}
\label{sec:results_guo}

\begin{figure}
\centering
\includegraphics[width=\linewidth]{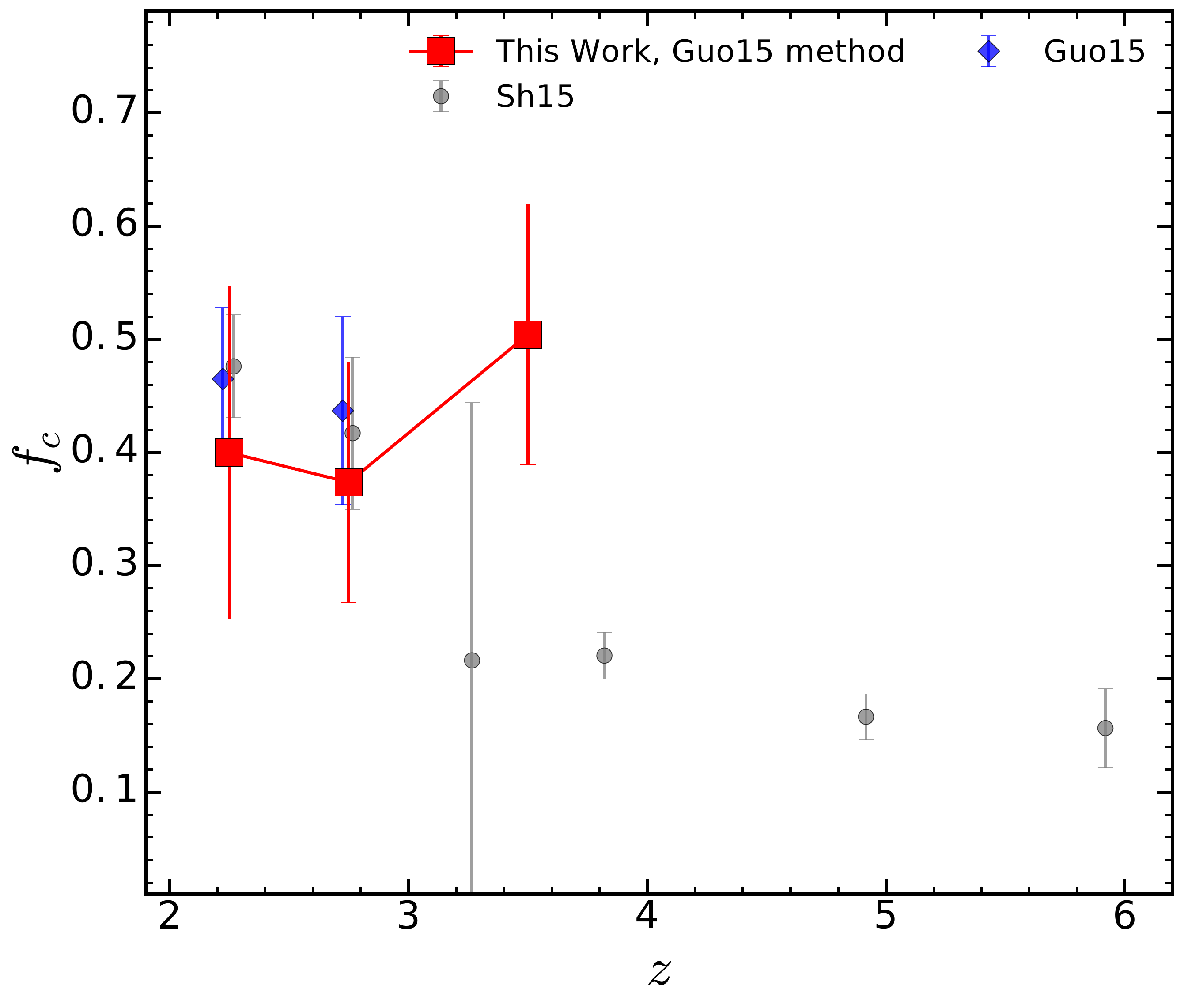}
\caption{Clumpy fraction in the VUDS sample as computed using the method described by \citet{guo2015} (red squares) and the results by \citet{guo2015} as blue diamonds and \citet{shibuya2016} as gray circles which use the same methodology.}
\label{fig:clumpy_fraction_guomethod}
\end{figure}

We compared the results obtained with our method to results obtained using the \citet{guo2015} method as defined in Sect. \ref{sec:guo_method} (see Fig. \ref{fig:clumpy_fraction_guomethod}). We find that for $z<3$ we are in excellent agreement with the results from \citet{guo2015} and \citet{shibuya2016} when we used a method similar to theirs. At higher redshift we included our clumpy fraction measurement  binned at $3<z<4$ to minimize statistical errors. We find that our clumpy fraction is higher than what is reported by \citet{shibuya2016} but it is compatible within $\sim2\sigma$ of our reported measurements. We thus conclude based on the bins for which we have larger samples hence more robust statistics that the main difference between our results and those reported by \citet{guo2015} and \citet{shibuya2016} is due to the different methods used to detect clumps in high redshift galaxies.

\section{Physical properties of clumps}\label{sec:clump_physprops}

\subsection{The relation between clump luminosity and area}

\begin{figure*}
\centering
\includegraphics[width=\linewidth]{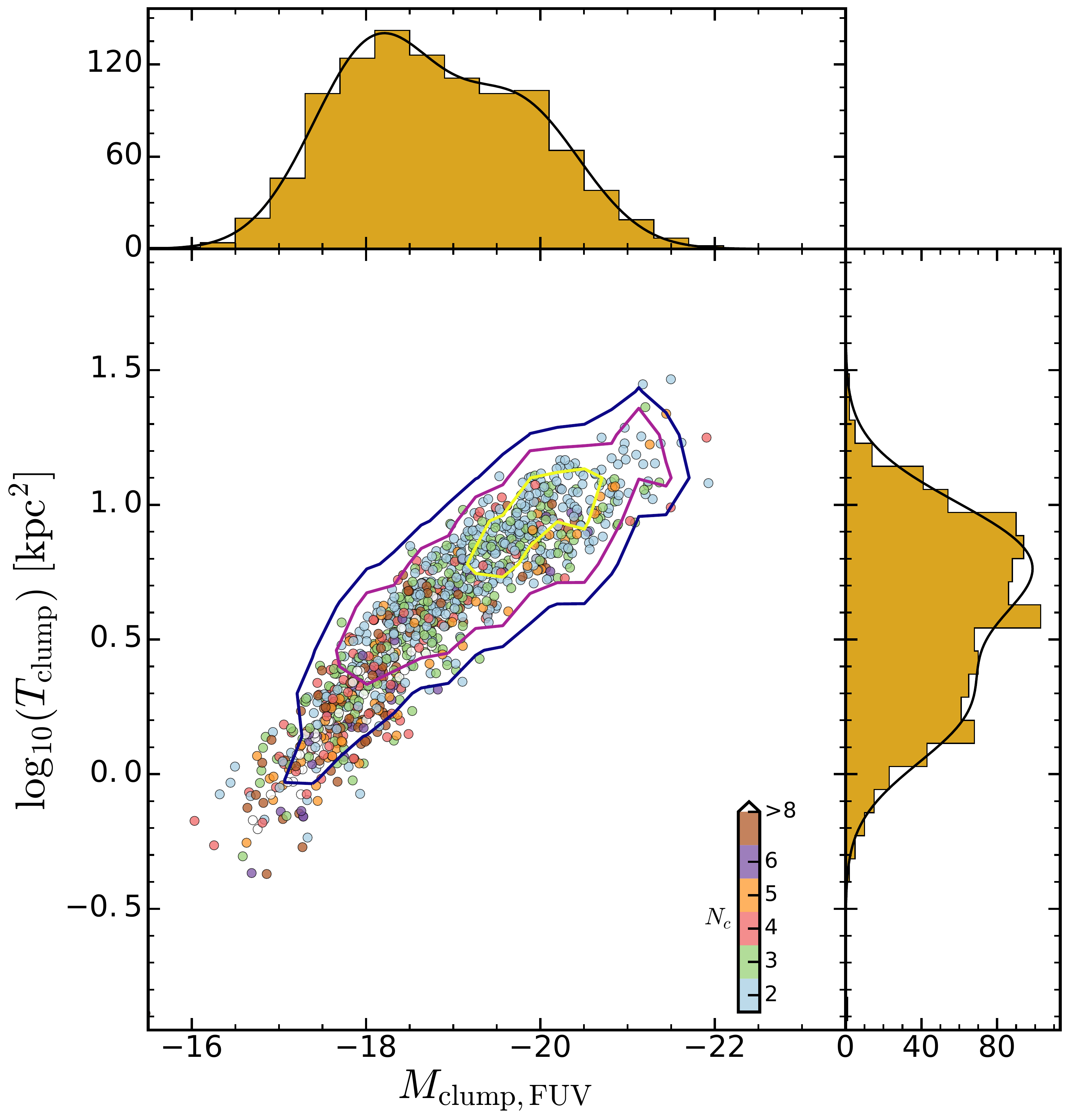}
\caption{Absolute magnitude in the rest-frame FUV and area distribution for clumps found in galaxies with $N_c\geq2$ detected at $2<z<4.5$ with $k_p=1.5$. Each clump is color coded by the number of clumps found in its parent galaxy. The color scheme is found at the bottom right of the main panel. The black solid line in each of the auxiliary panels shows a fit of a two Gaussian model to the observed distribution of areas and magnitudes of the clumps. We observe that it is more probable to find the brightest and largest clumps in two-clump systems. The contours show the smoothed distribution of the fraction of two-clump systems at levels of 10, 25, and 50\%. It is clear that the majority of two-clump galaxies (with respect to all clumpy systems) is found in brighter and larger clumps.}
\label{fig:clumps_physics_2}
\end{figure*}   

\begin{figure}
\centering
\includegraphics[width=\linewidth]{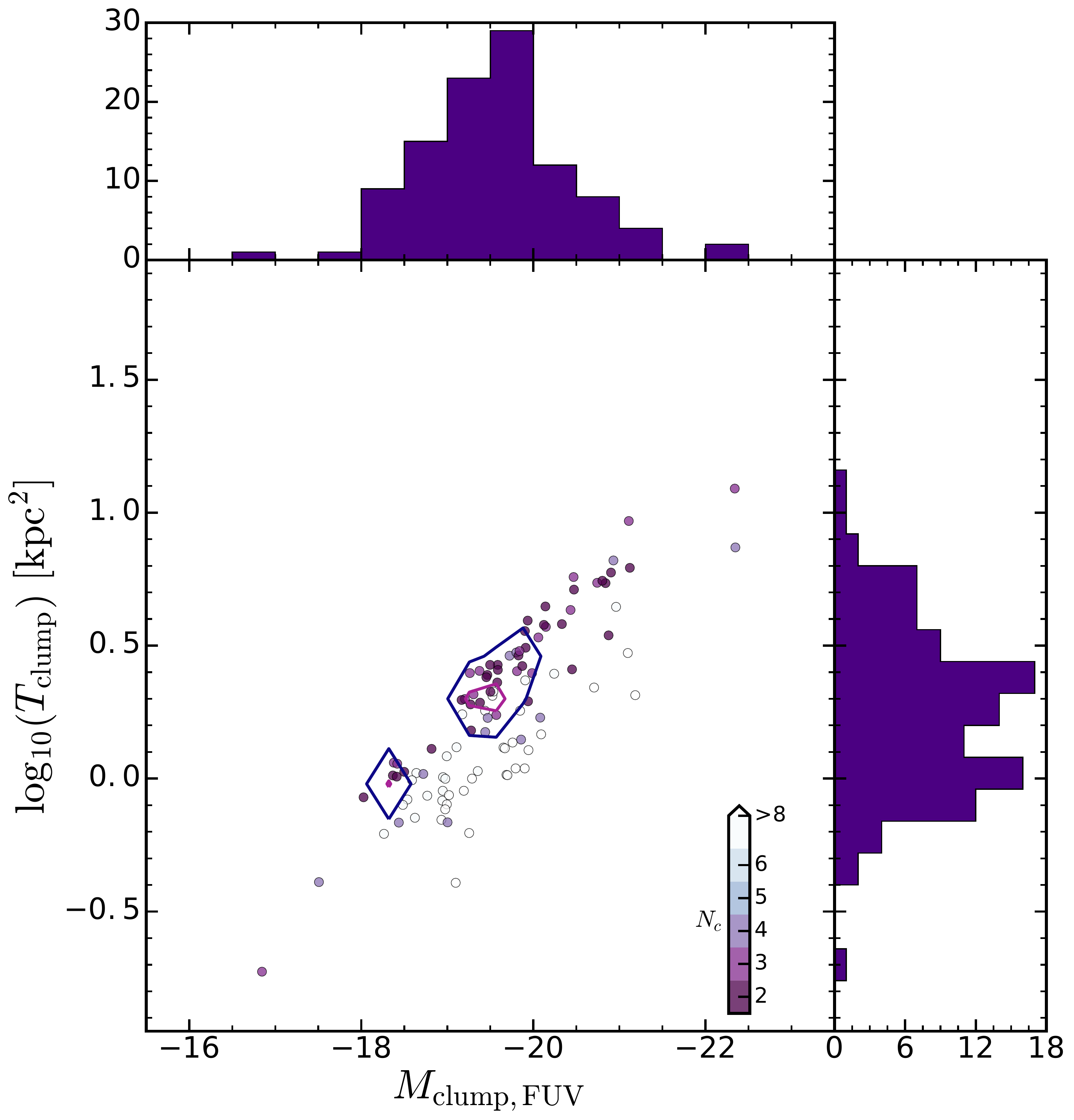}
\caption{Absolute magnitude in rest-frame FUV and area distribution for clumps found in galaxies with $N_c\geq2$ detected at $4.5<z<6.0$ with $k_p=5$. Each clump is color coded by the number of clumps found in its parent galaxy. The color scheme is found at the bottom right of the main panel. The contours shows the distribution of fraction of two-clump systems at levels of 10 and 25\%. }
\label{fig:clumps_physics_high}
\end{figure}

As stated in Sect. \ref{sec:clumps_physic}, we derived areas and absolute magnitudes for each clump by computing the flux and area within each group of pixels associated with the local maximum that defines the clump. We show in Figs.\ref{fig:clumps_physics_2} and \ref{fig:clumps_physics_high} the tight correlation that we observe between clump magnitude and clump area. This is expected by construction since the larger the area (i.e., the number of pixels of the clump) the more likely it is for it to be brighter. In this figure we exclude galaxies with a single clump detection.

We can observe in Fig. \ref{fig:clumps_physics_2} that the distribution of clump areas is quite broad, and that the magnitude distribution has an asymmetric shape. We fitted each of these distributions with a two Gaussian model with six free parameters: two amplitudes, two centers and two widths, one for each Gaussian curve. The fit was done using the Levenberg-Marquardt minimization algorithm implemented in \emph{scipy}. For clumps at $2<z<4.5$ we observe two peaks in the magnitude distribution at $M_{NUV}=-18.1\pm0.1$ and $M_{NUV}=-19.7\pm0.1$. In the area distribution the two peaks are found at $\log(T)=0.25\pm0.05$ and $\log(T)=0.78\pm0.03$, which correspond to $\sim1.8$ and $\sim6.0\ \mathrm{kpc}^{2}$. We test the statistical robustness of a possible bimodality using the Hartigan's dip test \citep{hartigan1985}. We find that the unimodal distribution hypothesis is highly favored (at $\sim80\%$ level). We do find a deviation from a normal distribution computed from the \citet{shapiro1965} test that rejects the normal distribution hypothesis at a 99\% level. At $4.5<z<6.0$ the low number of clumps does not allow us to test for a bimodal distribution (see Fig. \ref{fig:clumps_physics_high}).

We conclude from this analysis that a broad range of clump luminosities and sizes is observed. Clumps are confidently identified with areas as small as 1 kpc$^2$ and faint luminosities  M$_{clump, FUV}\sim -17$, or with sizes as large as 15 kpc$^2$ and bright luminosities M$_{clump, FUV}\sim -21$. This diversity in clump properties may be  linked to different processes of clump formation producing bright and large clumps preferentially with with M$_{clump, FUV}<-18.5$ and area T$_{clump}>4$ kpc$^2$, or faint and small clumps with M$_{clump, FUV}>-18.5$ and area T$_{clump}<4$ kpc$^2$, producing the continuous range of observed properties. This is discussed in Sect. \ref{sec:discussion}.

\subsection{Bright and large clumps are in two-clump galaxies}

A key piece of information shown in Fig. \ref{fig:clumps_physics_2} is the number of clumps in the galaxy that each clump belongs to. The larger and brightest clumps are most likely to be found in two-clump galaxies and as we move toward fainter and smaller clumps we observe that they are more commonly found in galaxies with a larger number of clumps. This information is summarized in the contour lines of  Fig. \ref{fig:clumps_physics_2} showing the distribution of the fraction of clumps in two-clump galaxies as a function of clump area and absolute magnitude. More than 50\% of the brightest and largest clumps are found in two-clump galaxies  and the probability of finding two clumps in a galaxy decreases for fainter clump magnitudes and smaller areas (down to $\sim10\%$). The information on how the fraction of clumps in two-clump galaxies evolves with redshift is found in Table \ref{tab:fractions}.

\subsection{Luminosity ratios of two-clump galaxies}

To classify each two-clump galaxy we compute the luminosity ratio between the two  clumps, $L_f/L_b$, with $L_f$ being the luminosity of the fainter clump and $L_b$ that of the brighter of the two. The resulting fractions as a function of redshift can be found in Table \ref{tab:fractions}. The fraction of two-clump galaxies is classified in major ($1/4<L_f/L_b<1$) and minor ($L_f/L_b<1/4$) systems, for galaxies which have at least one clump with $M_\mathrm{clump}<-19.1$. Statistical uncertainties become large because of the reduced number of galaxies for which we are able to compute this quantity.

\begin{table*}
\centering
\begin{tabular}{|c|cccc|c|}
\hline
redshift & $f_\mathrm{2c}\ [\%]$ & $f_\mathrm{2c,bright}\ [\%]$ & $f_\mathrm{major,bright}\ [\%]$ &  $f_\mathrm{minor,bright}\ [\%]$ &  $f_\mathrm{gal,pair}\ [\%]$\\
\hline
$2.0<z<2.75$ & $40\pm6.1$ & $60\pm15.4$ & $77\pm22.8$ & $23\pm9.9$ & $15\pm4.5$ \\
$2.75<z<3.5$ & $38\pm3.9$ & $60\pm9.0$ & $64\pm14.0$ & $36\pm9.4$ & $21\pm5.1$ \\
$3.5<z<4.5$ & $19\pm2.4$ & $37\pm6.3$ & $43\pm13.8$ & $57\pm17.0$ & $39\pm11.6$ \\
$4.5<z<6.0$ & $12\pm2.5$ & $15\pm3.4$ & $53\pm32.6$ & $47\pm29.7$ & $38\pm24.0$ \\
\hline
\end{tabular}
\caption{Fraction of two-clump systems in four redshift intervals at $k_p=1.5$ for $2<z<4.5$ and $k_p=5.0$ for $4.5<z<6$. The second column represents the total fraction of two-clump galaxies. The third column represents the fraction of clumps in two-clump galaxies with at least one of them being brighter than $M_\mathrm{clump}<-19.1$. The fourth and fifth columns represent the fraction of bright  systems that are in major ($1/4<L_f/L_b<1$) and minor ($L_f/L_b<1/4$) systems. The luminosities are measured in FUV rest-frame. The last column shows the pair fraction corrected for projection effects (see Sect. \ref{sec:result_clump}) considering pairs of $i_\mathrm{AB}<24.5$  VUDS targets with fainter galaxies.}
\label{tab:fractions}
\end{table*}

\subsection{Clump stellar masses}

We compute the stellar mass of clumps with a simple assumption that the mass to light ratio of clumps is the same as its host galaxy. This hypothesis allows us to estimate the stellar mass of each clump by computing the clump to galaxy luminosity ratio and applying this ratio to compute the stellar mass of a clump from the total stellar mass of the galaxy. Clearly, this estimate is a first order approximation as stellar clumps are likely to be on average more star-forming than the galaxy as a whole if they are strongly star-forming regions,  and we might therefore underestimate the stellar masses of clumps.

\begin{figure}
\centering
  \includegraphics[width=\linewidth,angle=0]{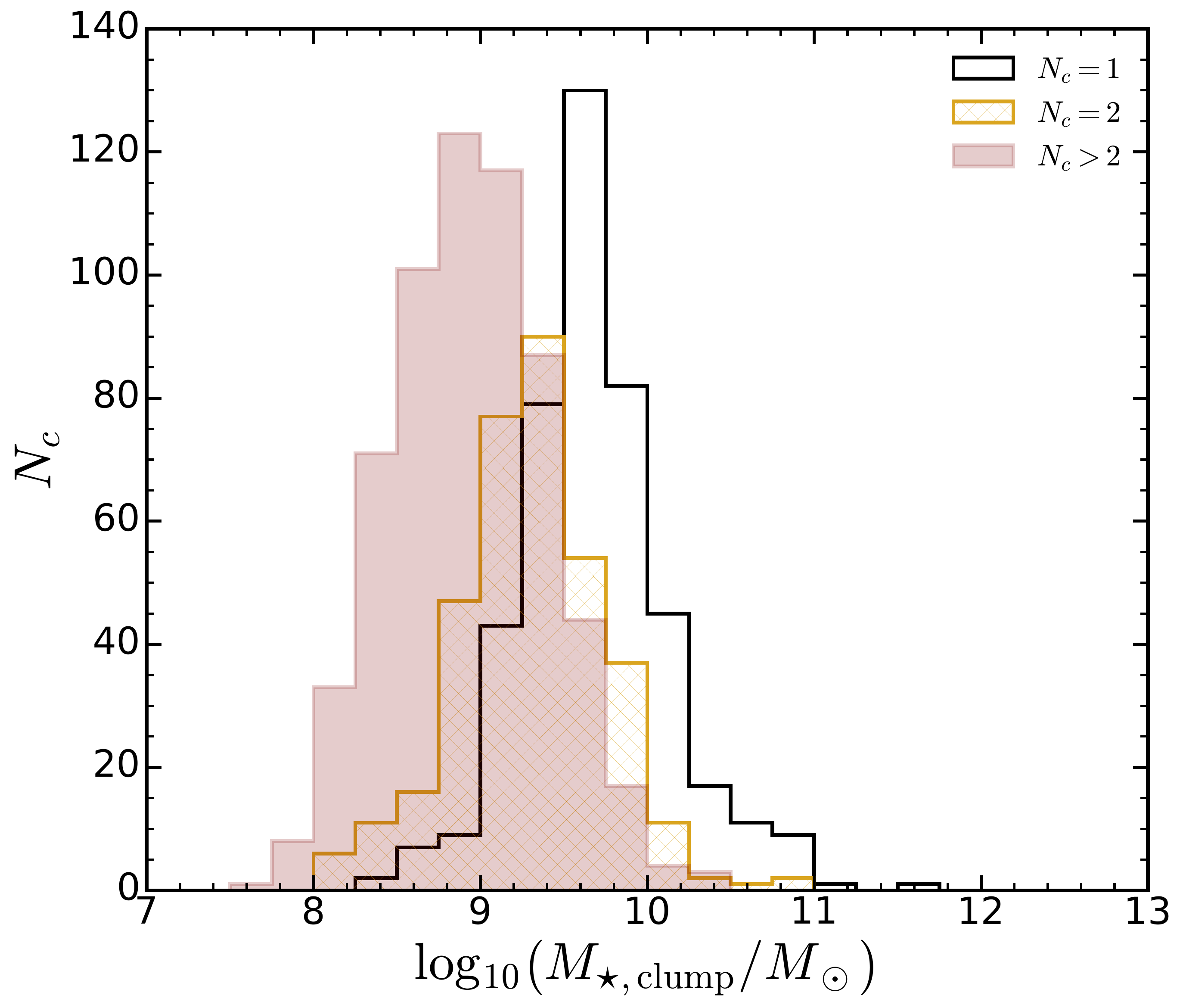}  \caption{ Clump stellar mass distributions for three galaxy subsamples. Black, gold, and filled pink histograms respectively indicate the distribution for galaxies with one clump, two clumps, and three or more clumps.}
  \label{fig:clump_mass}
\end{figure}

In Fig. \ref{fig:clump_mass} we compare the distribution of stellar masses in clumps for three different samples. A first sample is considering clumps that are in clump systems with at least three clumps (hereafter multi-clump systems). Another stellar mass distribution is for the sample of clumps in two-clump galaxies. These two samples are compared to the distribution of stellar masses in single clumps galaxies. We note that for single clump systems, a clump can correspond the entire galaxy, explaining why we find clumps with stellar masses up to $\log_{10}(M_\star/M_\odot)\sim11$.

From Fig. \ref{fig:clump_mass} it is striking to see that the clumps in two-clump galaxies have stellar masses that peak at higher values (around $\log_{10}(M_\star/M_\odot)\sim9.4$) than  clumps in multi-clump systems. The stellar masses of these clumps appear to be larger than what is expected for giant clumps identified from galaxy disk simulations or observations \citep[see e.g.,][]{elmegreen2009,genzel2011,guo2012,bournaud2016}, with such clumps having lower stellar masses in the range $8\lesssim\log_{10}(M_\star/M_\odot)\lesssim9$. On the other hand, the bulk ($\sim55\%$) of the stellar mass distribution in multi-clump-systems is consistent with this lower mass range. We have tested the similarity between the distributions of stellar masses in two-clump galaxies and in multi-clump systems using a Kolmogorov-Smirnov test and the null hypothesis (that both distributions are the same) is discarded at the 95\% confidence level. This difference persists when considering higher values of $k_p$. We conclude that these two populations are genuinely different, and discuss the possible reasons for these differences in Sect. \ref{sec:discussion}.

\section{Discussion: a merger origin for the bright and large clumps in two-clump galaxies}
\label{sec:discussion}

Our study brings to light two important morphological properties of high redshift galaxies and their evolution from $z\sim6$ to $z\sim2$: the fraction of clumpy galaxies as well as the distribution of the number of clumps per galaxy. Both of these properties are a direct consequence of the evolutionary path followed by galaxies, and the physical processes driving this evolution at these early cosmic times.

We find that the fraction of clumpy galaxies in the star-forming population rises from $\sim34\pm8\%$ at $z\sim2.25$ to $54\pm9\%$ at $z\sim4$ for galaxies selected in the UV rest-frame with stellar masses $\log_{10}({M_\star/M_\odot}) > -0.204\times(z-4.5)+9.35$, and identifying clumps brighter than $M_\mathrm{clump,NUV}<-18$ and larger than $\sim2\ \mathrm{kpc}^2$.  Our method has the benefit of producing counts that can be readily compared over a large redshift range, revealing that the clumpy fraction remains high over the redshift range $2<z<6$.

These results on the evolution of the clumpy fraction are apparently at odds with the decreasing fraction of clumpy galaxies reported by \citet{guo2015} and \citet{shibuya2016}, but can be easily understood from the difference in the surface brightness limits used to identify clumps. The surface brightness limit evolving with redshift that we use seems to be the key difference with   \citet{guo2015} and \citet{shibuya2016}.  We made the choice to use an isophote evolving with redshift (eq.2), as motivated by the general luminosity evolution with redshift.  In comparison, \citet{guo2015} and \citet{shibuya2016} use a fixed isophote level, independent of redshift. This implies that we use smaller search areas at the lowest redshifts (as isophote levels are higher in our method) and higher search areas at the highest redshifts (lower isophote levels in our method). This implies in turn that we count less clumps at low redshifts, and more clumps at high redshifts.

Other more subtle differences in the definition of a clumpy galaxy and in the sample used to compute the clumpy fraction may further contribute to the differences observed between our study and others. We define a clumpy galaxy as a galaxy with two or more detected clumps within its limiting isophote, whereas \citet{guo2015} and \citet{shibuya2016} define it when at least one off-center clump is identified, and we should then expect a lower clumpy fraction from our measurements. Another important difference resides in our sample selection which does not apriori exclude galaxies because of their morphology \citep[][impose an effective radius $r_e>0.2\arcsec$ and axis ratio $q>0.5$ cut]{guo2015,shibuya2016}. As we included in our sample small compact sources, this might further increase the number of non-clumpy galaxies and decrease the clumpy fraction, while including galaxies  with low values of $q$ like elongated clump chain galaxies \citep[see e.g.,][]{elmegreen2004,elmegreen2008} would rather increase it.

We therefore argue that the overall lowering of the clump fraction with increasing redshift as reported by previous studies may simply be due to the bias produced by using the same observed surface brightness at all redshifts when identifying clumps: because of the limits in detection at the highest redshifts one probes only the brighter and rarer clumps, while at the same observed limiting surface brightness one finds more numerous fainter clumps, hence mimicking a clump fraction decrease with redshift.   This important methodological difference leads to results  significantly different from previous literature results, and we believe that our method improves on earlier attempts with a more physically motivated method. 
Our report of a constant or slightly increasing clump fraction over $2<z<6$ is consistent with the large range of total sizes of galaxies that we report in a previous work \citep{ribeiro2016}, and indicates that the morphology of galaxies remains varied and complex from the end of reionization to the peak in star formation activity.

The other particularly important result from our paper is from the analysis of the distribution of the number of clumps per galaxy, were we observe a large fraction ($\sim21-25$\%) of galaxies with only two clumps. The clumps in this fraction of the galaxies are the brightest and largest as shown in Figs. \ref{fig:clumpy_examples2}, \ref{fig:clumps_in_vuds}, \ref{fig:clumps_physics_2}, and \ref{fig:clumps_physics_high}. We can identify only a few possibilities to form the clumps in these two-clump galaxies: they could be either (1) the result of disk fragmentation and VDI, (2) two individual galaxies in the final merging process of a galaxy pair,  or (3) the chance projection of a galaxy on the line of sight, physically unrelated. We note that we are able to detect clumps that are one to two magnitudes fainter than what we find for bright clumps in two-clump systems. This implies that if additional clumps were to be found with deeper imaging they would be much fainter than what we find and thus we would still observe a large fraction of two-clump systems detected above a given luminosity limit.

Galaxies with VDI-induced clumps have been studied in detailed hydrodynamical simulations which show how fragmentation in a proto-galactic disk produces several clumps, visible over timescales $\lesssim1$ Gyr, before migrating and merging in a galaxy central core \citep[e.g.,][]{bournaud2007,dekel2009,genel2012,bournaud2014,oklopcic2016}. 
It seems unlikely that a physical scenario following disk fragmentation and VDI would preferentially produce only two bright clumps (F. Bournaud, private communication). Cloud fragmentation due to disk instabilities are more likely to give rise to a higher number of clumps, with the number of clumps reducing over the dynamical timescale of clump merging, as shown in simulations \citep{mandelker2014}. If VDI is the only process capable of producing clumps one would rather expect to observe more galaxies with three or more clumps than two-clump galaxies, contrary to what we find (Fig. \ref{fig:clumps_physics_error}). It is however possible that some of the two-clump galaxies that we observe are the end result of clumps formed through VDI after migration and merging of smaller clumps. The typical timescale for clump migration to a galaxy center is about 500 Myr \citep[e.g.,][]{schreiber2011,guo2012,bournaud2016} which is typically one fourth of the redshift interval we probe from $2<z<4.5$. However, the period on which we would observe two resolved clumps should be even shorter, further reducing the fraction of large and bright clumps if produced by VDI. Additionally, we would need to take into account the survival rate of large clumps for which some simulations suggest even shorter lifetimes \citep[e.g.,][]{genel2012}.  We therefore infer that the large fraction of two-clump galaxies with large and bright clumps cannot be produced by VDI processes alone.

Another interpretation is that these two-clump galaxies are direct evidence for on-going merger events. Major or minor mergers do indeed produce a bi-modal light distribution with each of the merging galaxy counted as a clump. Merging is observed to be a significant process with a high fraction of $\sim15-20$\% of galaxies involved in major mergers at redshifts $z\sim1$ and up to $z\sim4$ \citep[e.g.,][]{lefevre2000,deravel2009,lopez-sanjuan2013,tasca2014}. The bright clumps in two-clump galaxies and a luminosity ratio $L_f/L_b>1/4$ have absolute magnitudes $-19>M_{NUV}>-22$, producing together a total luminosity equivalent to that of the median of the star-forming population we are probing. Two-clump galaxies with one bright clump and $L_f/L_b<1/4$ may include minor merger events participating to the build-up of stellar mass along cosmic time. 

If we assume that clumps which have $L_f/L_b>1/4$ are representative of major galaxy mergers we find that $24-40\%$ of two-clump galaxies are then classified as major merger events, remaining roughly constant at $z\gtrsim3$. As two-clump galaxies represent $\sim25$\% of the total population, using the fraction of bright two-clump galaxies (Table \ref{tab:fractions}) we find a total merger fraction slightly decreasing with redshift from $\sim18$\% at $z\sim3$ to $\sim10$\% at $z\sim5$. Using those systems with $L_f/L_b>1/4$ we find a major merger fraction in the range 6.5 to 11\%. Similarly, using the fraction of two-clump galaxies with  $L_f/L_b<1/4$ we find a minor merger fraction of 2.7 to 10.5\%. 

The fraction of merging systems derived above is obtained when considering close merging pairs alone, hence late stage mergers, as the median clump separation is $\sim$3 kpc ($\sim$0.4 \arcsec). To expand this to a total merger fraction we need to account for the fraction of galaxy pairs within a 20 kpc search radius, a distance for which the probability of galaxies to merge in a short timescale is expected to be high \citep[e.g.,][]{Kitzbichler2008}. From a pair count corrected for projections along the line of sight (Sect. \ref{sec:result_clump}), we find a rising pair fraction of $\sim15\%$ at $z\sim2.3$ reaching $\sim38\%$ at $z\sim5$. One interesting trend is that while the pair fraction is increasing at higher redshifts, the opposite trend is observed for the merger fraction derived from clumps statistics ($f_\mathrm{2c}$ in Table \ref{tab:fractions}). One may  infer that we are observing an evolving merger rate measured in two stages, the first and more common at the higher redshifts is when galaxies have yet to merge, and the second, as observed from close bright clumps counts, representing the final phases of merging before coalescence, which is more common at lower redshifts. When combining these two merging indicators we find that the major merger rate remains roughly constant at $\sim20$\% over the redshift range $2<z<6$. 

There is a low probability that our observations of two-clump galaxies might be consequence of line of sight contamination from lower redshift galaxies which might then be confused with high redshift clumps. With a small median separation of  $\sim$0.4 \arcsec, the probability that these two-clump galaxies with at least one bright $M_\mathrm{clump,NUV}<-19.1$ are the result of random line of sight contamination  separations is very small, at less than 3\% (see Sect. \ref{sec:result_clump} for details), even when dealing with faint objects with $i_\mathrm{AB}\sim27-28$. We then conclude that most of these two-clump galaxies are real physical systems at the same redshift.

As a conclusion, a straightforward interpretation of galaxies with two bright clumps is that this population predominantly includes galaxies caught in the act of a merging event. This interpretation is further reinforced when looking at the typical stellar masses of clumps. We find that for those two-clump systems with at least one bright component, the majority ($\sim77\%$) have stellar masses greater than $\log(M_\star/M_\odot)>9$ a limit beyond which clumps induced by VDI are not observed in simulations \citep[see e.g.,][]{elmegreen2009,guo2012,tamburello2015,bournaud2016}. On the other hand, $\sim55\%$ of clumps in multi-clump systems have stellar masses lower than  $\log(M_\star/M_\odot)<9$, which might likely result from VDI as well as minor mergers. We find that the typical masses of the clumps in two-clump systems are more alike the masses of galaxies observed in major merging pairs with larger separations \citep{lopez-sanjuan2013,tasca2014}.

From the analysis of clump properties presented in this paper, we therefore conclude that we are likely witnessing  two different modes of clump formation happening in parallel at the same cosmic times. Small and faint clumps are likely the result of both VDI clumps production and minor mergers. At the same cosmic time, we do observe an important fraction (in $\sim20-25\%$ of all galaxies) of two-clump galaxies with large and bright clumps which are most likely major mergers. 
 
While our conclusions are based on indirect evidence, direct observational evidence will be required to confirm the physical origin of the two populations of clumps that we have identified. The definite proof that the bright two-clump galaxies are indeed major merging systems would require dynamical information on the velocity field, with the relative velocity separation of the two bright clumps. Some of these merging systems are already confirmed from spectroscopic evidence in the VUDS survey \citep{tasca2014}, and more evidence of a high merger fraction will be presented in a forthcoming paper (Le F\`evre et al., in prep). Likewise, velocity maps of multiple clump systems could bring some more evidence in support of clump formation via instabilities, although the dynamical signature of a VDI process may be rather subtle \citep{bournaud2009} and may escape current or even upcoming observational capabilities. Only the result of the VDI process in the form of a clump or several clumps, as shown in simulations, is accessible to observations today as reported here. Identifying minor mergers within the population of small and faint clumps is also important but again will require significantly improved observational capabilities. Further work is therefore needed to consolidate the picture we propose, but our results point out the diversity of clump properties, which signals a likely diversity in their formation process.

\section{Summary}
\label{sec:summary}

The results presented in this paper can be summarized as follows:
\begin{itemize}
\item The search for clumps in distant galaxies  images critically depends on the depth of the images defining the search area, on how clumps are defined, whether clumps must be embedded within a common isophote, and on the selection of the galaxy population. We used a new algorithm for detecting clumps, in which clumps are defined as groups of connected pixels associated with the same local maximum within a galaxy isophote evolving with redshift following surface brightness dimming and luminosity evolution,  and corrected for color differences between the different rest-frame bands observed as a function of redshift.
\item We computed the number of clumps per galaxy in a sample of 1242 galaxies with $2<z<4.5$ and 96 galaxies with $4.5<z<6$ with confirmed spectroscopic redshifts from the VIMOS Ultra Deep Survey and in the COSMOS and ECDFS areas where deep HST imaging in the F814W filter is available from the COSMOS and CANDELS surveys. 
\item The fraction of clumpy galaxies, with two or more clumps and corrected for line of sight chance projections, is found to remain in the range 35 to 55\% over $2<z<6$,  possibly increasing  with redshift.  
\item We find that at $2<z<4.5$ and excluding single clump galaxies, the dominant galaxy population has two clumps ($\sim25\%$ of all galaxies) followed by galaxies with three, and more than three clumps ($\sim11,\sim 7\%$ respectively). The same trend is found at higher redshifts ($4.5<z<6$) where galaxies with two clumps dominate ($\sim21\%$)  followed by galaxies with three, and more than three clumps ($\sim8,\sim 9\%$ respectively). 
\item We find a large range of clump properties, with a continuous, possibly bi-modal, distribution of the luminosity and area of clumps with a population of large and bright clumps distinct from a population of small and less luminous clumps. The population of large and bright clumps has typical absolute magnitudes $M_\mathrm{FUV}\sim-19.7$ and areas $T\sim6.0 \ \mathrm{kpc^{2}}$. In contrast, the population of small and less luminous clumps has typical absolute magnitudes $M_\mathrm{FUV}\sim-18.1$ and areas $T\sim1.8 \ \mathrm{kpc^{2}}$. We associate these two populations to the existence of two physical processes, mergers and disk instabilities, that explain the observed properties of clumps.
\item The faint and small clumps are preferentially found in multiple clump galaxies with N$_{clump}\geq$3. They are comparable in luminosity and stellar mass ($10^{8}-10^{9}M_\sun$) to the large star clusters seen forming in  numerical simulations from violent disk instabilities, consistent with findings from other studies \citep[e.g.,][]{elmegreen2009,guo2012,bournaud2016}. It is also likely that a fraction of these clumps is related to minor merging events. 
\item The bright and large clumps are preferentially found in two-clump galaxies. These bright clumps have luminosities and stellar masses $\sim 10^{9}-10^{10}M_\sun$, larger than expected from VDI processes. It appears as unlikely that, if clumps were solely produced by VDI, two-clump galaxies would strongly dominate the population of clumpy galaxies. We rather conclude that the bright and large clumps in two-clump systems have properties similar to those of galaxies in merging pairs \citep{lopez-sanjuan2013,tasca2014}. Using the fraction of galaxies with bright clumps we infer a major merger fraction going from 18\% at $z\sim3$ to 10\% at $z\sim5$. When combined with the pair fraction of galaxies as measured in projection within a 20 kpc radius, we find a major merger rate of $\sim20$\% comparable to other estimates at similar redshifts  \citep{lopez-sanjuan2013,tasca2014}.
\end{itemize}

In quantifying the properties of clumps in star-forming galaxies with $2<z<6$ we therefore conclude that we are witnessing the effects of two different galaxy assembly processes working in parallel at a cosmic time of major galaxy assembly. Clumps are likely to be the result of both violent disk instabilities with an in-situ clump formation mode, as well as  from major and minor merging of galaxies assembling matter coming from different origins. On the one hand gas accretion is expected to feed the buildup of disks which then become unstable and produce clumps that will build galaxy bulges \citep[e.g.,][]{dekel2009,bournaud2009,dekel2013,bournaud2016}, a process which remains to be confirmed by further observational evidence at these high redshifts \citep[e.g.,][]{bouche2013}. The  population of large and bright merging clumps formed ex-situ from the galaxy they are assembling into confirm mounting evidence for the importance of major mergers in the early build-up of galaxies, events which contribute the most to the stellar mass growth of a galaxy \citep{lopez-sanjuan2013,tasca2014,mandelker2014}.

Both disk formation processes and early merging events are therefore important to understand both the evolution of the cosmic star formation rate and the build-up of stellar mass of galaxies throughout cosmic time, and explaining key observables such as the evolution of the specific star formation rate, and its possible flattening at redshifts $z>2$ \citep[e.g.,][]{dekelm2014,tasca2015,faisst2016,marmol2016}.
We point out that in this study we did not discuss the properties of galaxies with a single bright clump, which dominate the population of star-forming galaxies. This population will be the subject of forthcoming papers. 

Testing the evolution of the number of clumps from larger simulations of galaxies in complete cosmologically representative volumes would bring further insight into the relative contribution of these different physical processes in building-up galaxies along cosmic time.  
On the observational side, since the clumpy fraction and the number of clumps identified strongly depend on the spatial resolution and depth of images, and as galaxies with large sizes are observed up to large redshifts \citep{ribeiro2016}, it is likely that future facilities like JWST and extremely large telescopes with adaptive optics may well identify more clumps. Spatially resolved information from multi-band high resolution observations and/or IFU data will be necessary to compute velocity fields and stellar ages of the large and bright clumps and confirm or not a scenario where these are witnesses of the final stages of merging pairs of galaxies. The high fraction of bright two-clump galaxies and the wide range of clump properties reported here must be taken into account when planning for future surveys aimed at a complete census of the properties of the high redshift population.



\begin{acknowledgements}
We thank the anonymous referee for his/her comments and suggestions which helped improve this manuscript.s
This work is supported by funding from the European Research Council Advanced Grant ERC--2010--AdG--268107--EARLY and by INAF Grants PRIN 2010, PRIN 2012 and PICS 2013. R.A. acknowledges support from the ERC Advanced Grant 695671 ‘QUENCH'. 
This work is based on data products made available at the CESAM data center, Laboratoire d'Astrophysique de Marseille. 
BR gratefully thanks Ana Afonso for late-night and helpful discussions, style tips and English typo corrections.
This research made use of Astropy, a community-developed core Python package for Astronomy \citep{astropy2013}; Numpy \& Scipy \citep{numpy_scipy}; Matplotlib \citep{Hunter:2007}.

\end{acknowledgements}


\bibliographystyle{aa}
\bibliography{refs_clumps}


\end{document}